\documentclass{aa}  

\usepackage{graphicx}
\usepackage{txfonts}
\newcommand{\nodata}{...}
\begin{document}

   \title{The first detection of ultra-diffuse galaxies in the Hydra I cluster from VEGAS survey}

   \author{E. Iodice\inst{1,2}
          \and
          M. Cantiello\inst{3}
          \and
          M. Hilker\inst{2}
          \and
          M. Rejkuba\inst{2}
          \and
          M. Arnaboldi\inst{2}
          \and
          M. Spavone\inst{1}
          \and
          L. Greggio\inst{4}
          \and
          D. A. Forbes\inst{5}
          \and
          G. D'Ago\inst{6}
          \and
          S. Mieske\inst{7}
          \and
          C. Spiniello\inst{1,8}
          \and
          A. La Marca\inst{9}
          \and
          R. Rampazzo\inst{10}
          \and
          M. Paolillo\inst{1,9,11}
          \and
          M. Capaccioli\inst{1,9}
          \and
          P. Schipani\inst{1}
          }

   \institute{INAF $-$ Astronomical Observatory of Capodimonte, Salita Moiariello 16, I-80131, Naples, Italy\\
              \email{enrichetta.iodice@inaf.it}
              \and
              European Southern Observatory, Karl-Schwarzschild-Strasse 2, D-85748 Garching bei Muenchen, Germany
              \and
             INAF-Astronomical Abruzzo Observatory, Via Maggini, 64100, Teramo, Italy
              \and
              INAF $-$ Osservatorio Astronomico di Padova, Vicolo dell'Osservatorio 5, I-35122 Padova, Italy
              \and
              Centre for Astrophysics and Supercomputing, Swinburne University of Technology, Hawthorn, Victoria 3122, Australia 
              \and
              Instituto de Astrofísica, Facultad de Física, Pontificia Universidad Católica de Chile, Av. Vicu\~{n}a Mackenna 4860, 7820436 Macul, Santiago, Chile 
             \and
             European Southern Observatory, Alonso de Cordova 3107, Vitacura, Santiago, Chile
             \and
             Department of Physics, University of Oxford, Denys Wilkinson Building, Keble Road, Oxford OX1 3RH, UK
             \and
             University of Naples ``Federico II'', C.U. Monte Sant'Angelo, Via Cinthia, 80126, Naples, Italy
             \and
              INAF $-$ Astronomical Observatory of Padova, Via dell’Osservatorio 8 , I-36012, Asiago (VI), Italy
              \and
             INFN, Sezione di Napoli, Napoli 80126, Italy
}

   \date{Received ....; accepted ....}

 
  \abstract{In this paper we report on the discovery of {  27 low-surface brightness galaxies, of which} 
  12 are candidate ultra-diffuse galaxy (UDG) 
  in the Hydra I cluster, based on deep observations taken as part of the {\it VST Early-type Galaxy Survey} (VEGAS). This first sample of UDG candidates in the Hydra I cluster represents an important step in our project that aims to enlarge the number of confirmed UDGs and, through study of statistically relevant samples, constrain the nature and formation of UDGs. 
  This study presents the main properties of this class of galaxies in the Hydra I cluster.
  For all UDGs, we analyse the light and colour distribution, and provide a census of the globular cluster (GC) systems around them. 
  {  Given the limitations of a reliable GC selection based on two relatively close optical bands only, 
  we find that half of the UDG candidates have a total GC population consistent 
  with zero. 
  Of the other half, two galaxies have a total population larger than zero at 2$\sigma$ level.}
  We estimate the stellar mass, the total number of GCs and the GC specific frequency ($S_N$). Most  of  the candidates  span  a  range  of  stellar  masses  of  $10^7-10^8$~M$_{\odot}$. 
  {  Based on the GC population of these newly discovered UDGs, 
  we conclude that most of these galaxies have a standard  or low dark matter content,   with a halo mass of $\leq 10^{10}$~M$_{\odot}$. 
  }
  
}



\titlerunning{UDGs in Hydra I cluster}
\authorrunning{E. Iodice et al. }

   \maketitle
%
\section{Introduction}\label{sec:intro}
Ultra-diffuse galaxies (UDGs) are extreme low-surface brightness (LSB) objects 
($\mu_g \geq 24$ mag arcsec$^{-2}$) with effective radii comparable to that of large spirals, but stellar masses similar to dwarf galaxies ($\sim 10^7 - 10^8$ M$_{\odot}$). 
Renewed interest in the LSB objects and the definition of a new class of galaxies as UDGs comes from 
the discovery of a significant population of these rather extreme objets in the Coma cluster  \citep{vanDokkum2015,Koda2015}. 
Recent works report the discovery of UDGs also in less dense environments as groups of galaxies and in the field \citep{Roman2017a,vanderBurg2017,Shi2017,Muller2018,Prole2019a,Forbes2019,Forbes2020b}.

The nature and origin of the UDGs are still debated. 
UDGs could be “failed” galaxies, which lost their gas supply at early epochs. If so, the UDG should posses a massive dark matter (DM) halo to survive in dense environments like galaxy clusters \citep{vanDokkum2015}. 
Some theoretical models explain UDGs as extreme dwarf galaxies, whose large size could be due to high spins in DM halos
\citep{Amorisco2016,Rong2017,Tremmel2019} or to tidal interactions \citep[e.g.][]{Yozin2015}.
\citet{diCintio2017} proposed that very extended UDGs-like systems could have formed by the 
kinematical heating of their stars as consequence of internal processes 
(i.e. gas outflows associated with feedback).\\ 
Populations of UDGs have been {  reproduced in cosmological galaxy simulations both in galaxy clusters \citep[e.g.][]{Sales2020}} and in low-density environments \citep[e.g.][]{Wright2020}. 
They suggested that two classes of UDGs might exist: one, found in the field, defined as normal LSB galaxies, 
and a second class of UDGs, near cluster centres, which shaped their large size and low surface brightness 
by the cluster tidal forces. 

From the observational side, given the very low surface brightness, 
the detection and analysis of UDGs is challenging.
Some UDGs in Coma and Virgo clusters appear to host a large number of globular clusters (GCs) for their luminosity, 
 indicating that they may be DM-dominated. This might support the failed galaxies hypothesis {   \citep{Beasley2016a,PengLim2016, vanDokkum2016,vanDokkum2019,Forbes2020a}.} 
On the other hand, other works showed that UDGs can have stellar masses and DM content 
consistent with dwarf galaxies, suggesting that they could be   extremely extended dwarfs \citep{Beasley2016,Amorisco2018,Alabi2018,Ferre-Mateu2018}. 
Furthermore, some UDGs are found with very low DM content 
\citep{vanDokkum2016,vanDokkum2019b,Danieli2019,Prole2019b}.

From the analysis of the average colours, two populations of UDGs were identified: red and quenched UDGs, which occupy the red sequence 
\citep{vanDokkum2015,vanderBurg2017}, and a blue population of UDGs, which are mostly found in the field \citep[e.g.][]{Leisman2017,Roman2017a,Prole2019a}.
The spatial distribution of UDGs inside their host environment seems to be asymmetric, i.e. 
over-densities of UDGs are found close to substructures, as sub-groups of galaxies, which are 
falling into the cluster \citep{vanderBurg2017,Janssens2019}.

The few spectroscopic data acquired for UDGs find both  metal 
poor ($ -0.5 \leq [M/H] \leq -1.5$) and old  systems \citep[$t\sim9$~Gyr, e.g.][]{Fensch2019,Ferre-Mateu2018,Gu2018,Ruiz-Lara2018}, as well as  
younger UDGs with an extended star formation history and massive DM halo \citep{Martin-Navarro2019}.

%
%
The available data show that UDGs span a wide range of properties, which do not fit in a unique formation scenario, and clearly indicate that, to date, our knowledge of UDGs is still poorly constrained. The main issues that need to be tackled are 
{\it i)} whether or not two classes of UDGs exist, one as extreme case of dwarf galaxies 
and the other as pure UDGs, with distinct formation scenarios, structural parameters and photometric properties; 
and {\it ii)} what is the fraction of baryonic versus DM content in UDGs, and whether this fraction differes 
for the two putative classes. 
if the majority of UDGS turns out to be without DM, then the current paradigm of galaxy formation should be revised and better understood in the specific parameter space occupied these galaxy systems.

In this letter we report the discovery of 12 UDG candidates in the Hydra I cluster. 
This is a rich environment of galaxies located at $\sim 51$~Mpc 
\citep{Christlein2003}, with a virial mass of $\sim 2 \times 10^{14}$~M$_{\odot}$ \citep{Girardi1998}, from
which we derived a virial radius $R_{vir} \sim1.6$~Mpc.
The cluster core is dominated by the two brightest early-type galaxies, NGC~3309 and NGC~3311, 
embedded in an extended diffuse stellar halo \citep{Arnaboldi2012}.
Most of the recent studies of Hydra I focused on the light distribution and 
kinematics in the cluster core, which showed the presence of ongoing interactions and traced  the 
extended mass assembly around NGC~3311 \citep{Ventimiglia2011,Richtler2011,Coccato2011,Misgeld2008,Misgeld2011,Arnaboldi2012,Koch2012,Barbosa2018,Hilker2018}.
We carry out here the first search for UDGs enabled by a new wide area and deep imaging dataset.




\section{Observations, data analysis and UDGs identification}\label{sec:data}

The Hydra I cluster is a target of the {\it VST Early-type Galaxy Survey} (VEGAS\footnote{see \url{http://www. na.astro.it/vegas/VEGAS/Welcome.html}}), a multi-band ({\it u,g,r,i}), deep imaging survey carried out 
with the European Southern Observatory (ESO) VLT Survey Telescope (VST). 
VST is a 2.6~m wide field optical telescope \citep{Schipani2012}, equipped with OmegaCAM, a $1^{\circ} \times 1^{\circ}$ camera with a resolution of $0.21$~arcsec~pixel$^{-1}$. 
The imaging data for the Hydra~I cluster, presented in this work, were collected in the $g$ and $r$ bands, 
in dark time, with a total integration times of 2.8 and 3.22 hours, respectively.
The measured image quality had an average FWHM$\sim 0.8$~arcsec.  
Data were acquired with the {\it step dither} observing strategy, consisting of a cycle 
of short exposures ($\sim$~$150$~sec) on the science target and on an adjacent field (close in space and time) to the 
science frame. This strategy was adopted also for other VEGAS targets \citep{Spavone2018,Iodice2020} 
and for the Fornax Deep Survey \citep[FDS,][]{Iodice2016,Venhola2018,Iodice2019}. 
It certifies a very accurate estimate of the sky background. 
For the Hydra~I dataset, a 7-magnitude bright star falls on the NE side of the cluster core, and was always put in one of the two wide OmegaCAM gaps during observations, to reduce the scattered light.

The data reduction was performed using {\it VST-Tube}, which is one of the dedicated pipelines to process 
OmegaCAM observations \citep{grado12,Capaccioli2015}. 
For the final stacked images we estimated surface brightness depths\footnote{Derived as the flux corresponding to $5 \sigma$, with $\sigma$ estimated over an empty area of 1 arcsec.} 
of  $\mu_g=28.6\pm0.2$~mag and $\mu_r=28.1\pm0.2$~mag
in the $g$ and $r$ bands, respectively. 
The light of the two bright stars close to the cluster core 
has been modelled and subtracted from the reduced images.

\begin{figure*}
	\centering
	\includegraphics[width=\hsize]{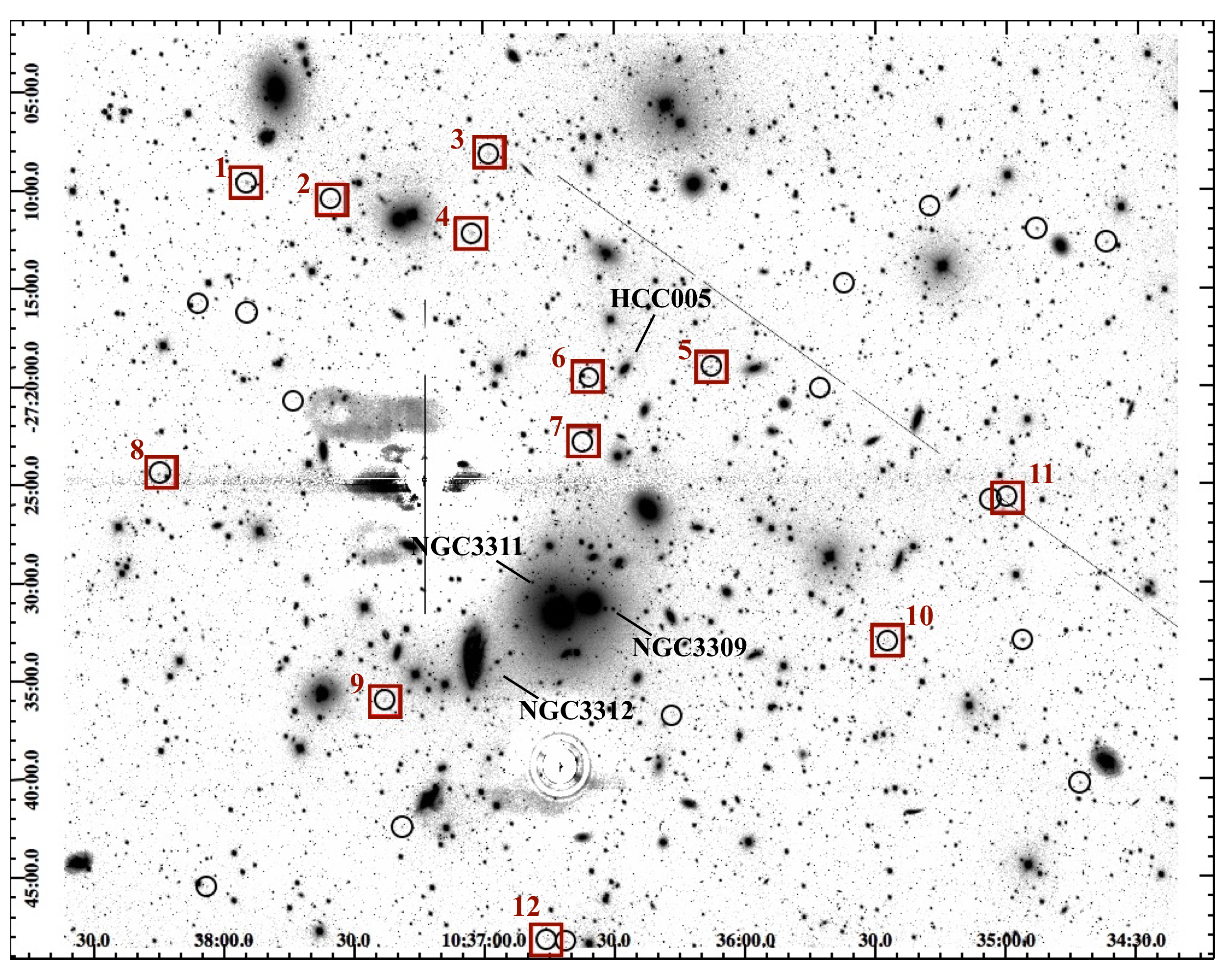}
	\caption{VST mosaic ($56.7 \times 46.55$~\arcmin $\sim0.8 \times 0.7$~Mpc) of the Hydra I cluster 
	in the $g$ band, which correspond to $\sim0.5 R_{vir}$. NGC~3311 and NGC~3309 are the two brightest cluster galaxies located close to the image centre, embedded in the extended diffuse light envelope. The 27 LSB galaxies detected in this work are marked as black circles. Red squares show the UDGs.
	North is up and East is on the left.}
	\label{fig:mosaic}
\end{figure*}

The final VST mosaic extends over $1^{\circ} \times 2^{\circ}$ ($0.9\times1.8$ Mpc), 
which covers the cluster out to the virial radius ($\sim 1.6$~Mpc). 
A portion of the mosaic centred on the cluster core is shown in Fig.~\ref{fig:mosaic}.
By visual inspection of this area, looking for faint, diffuse and extended objects, 
we identified 27 LSB galaxies that are not included 
in previous catalogues \citep{Christlein2003,Misgeld2008}, since they are below the detection limits of such studies
(see Sec.~\ref{sec:results}).
All 27 identified LSB candidates are marked with black circles in Fig.~\ref{fig:mosaic}.  

For each LSB candidate we 
{\it i)} extracted a thumbnail from the mosaic (three times more extended than the target) 
where all bright stars are modelled 
and subtracted, while the remaining fainter sources are masked, and a residual local background is 
 subtracted; 
{\it ii)} performed the isophote fitting, by using the IRAF task ELLIPSE, 
to obtain the azimuthally averaged $g$ and $r$-band surface brightness profiles, the total magnitudes and the average colours; 
{\it iii)} derived the 2-dimensional (2D) fit of the galaxy light, in the $g$ band, by using GALFIT  \citep{Peng2010} and adopting a single Sersic function, with all structural parameters left free.
All LSB galaxies are on average 10 to 100 times more extended than the FWHM of the point-spread function, therefore
we discarded the convolution step in the modelling and simply excluded the central regions affected by the seeing disk ( $\leq 1$~arcsec).
%

\begin{figure*}
	\centering
	\includegraphics[width=6cm]{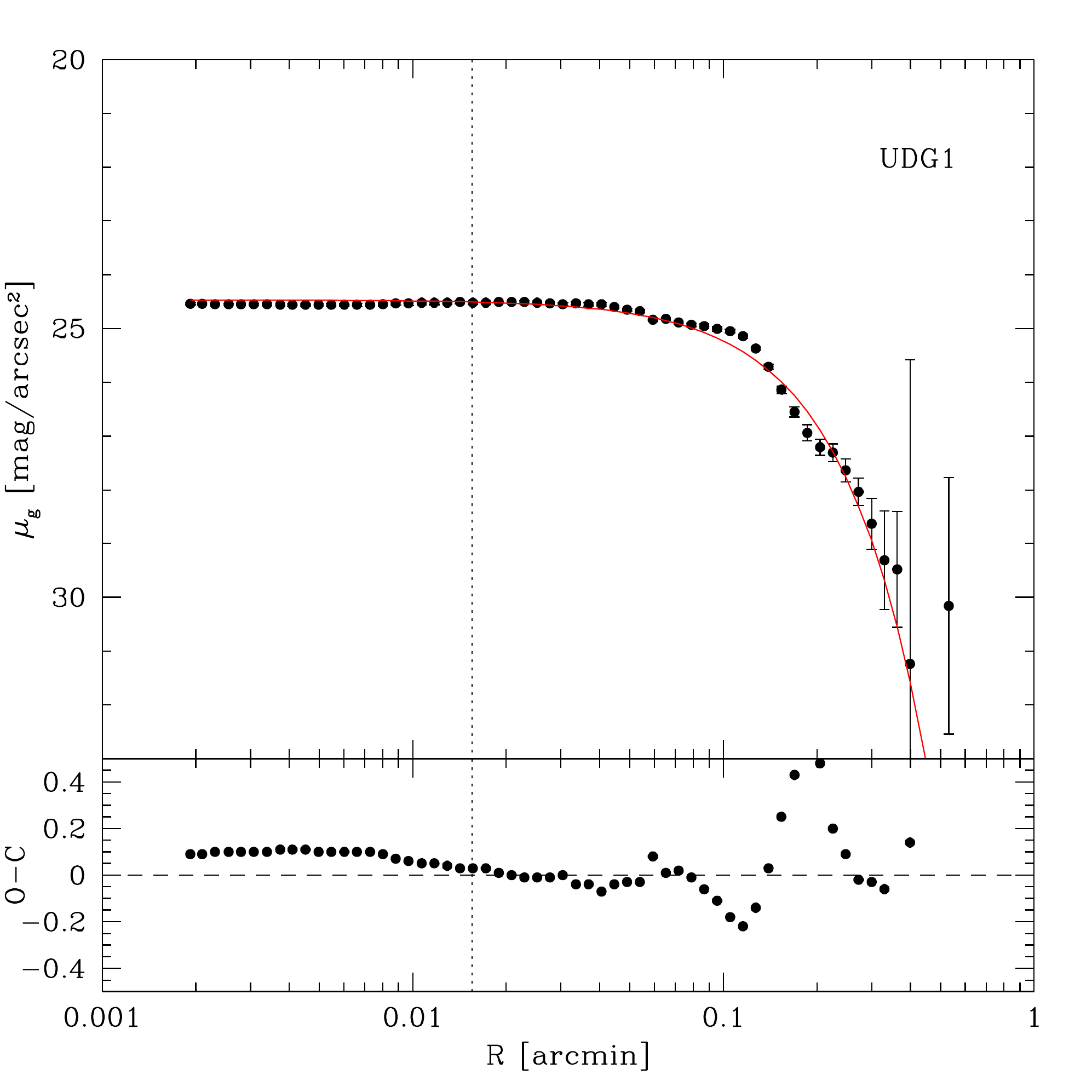}
	\includegraphics[width=6cm]{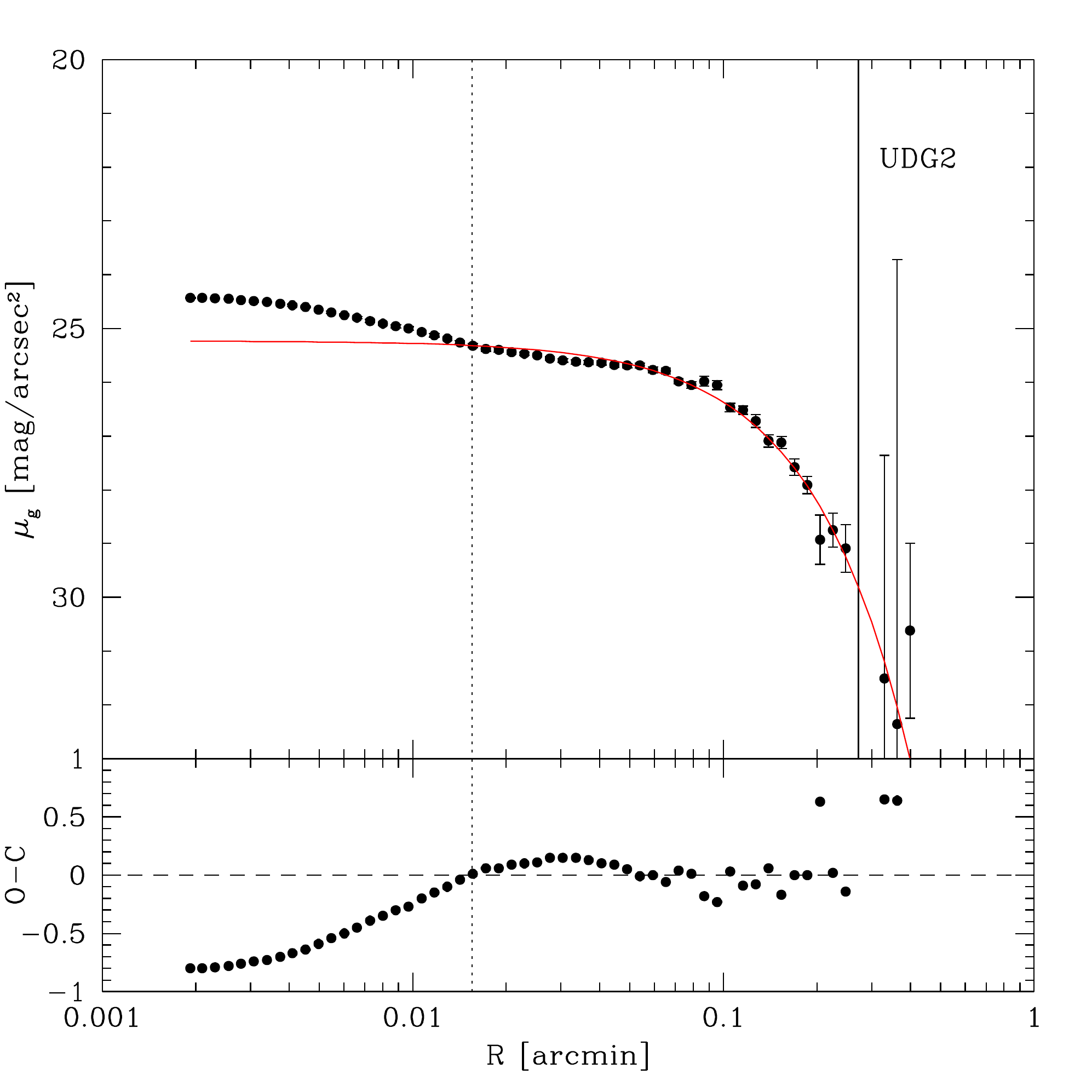}
	\includegraphics[width=6cm]{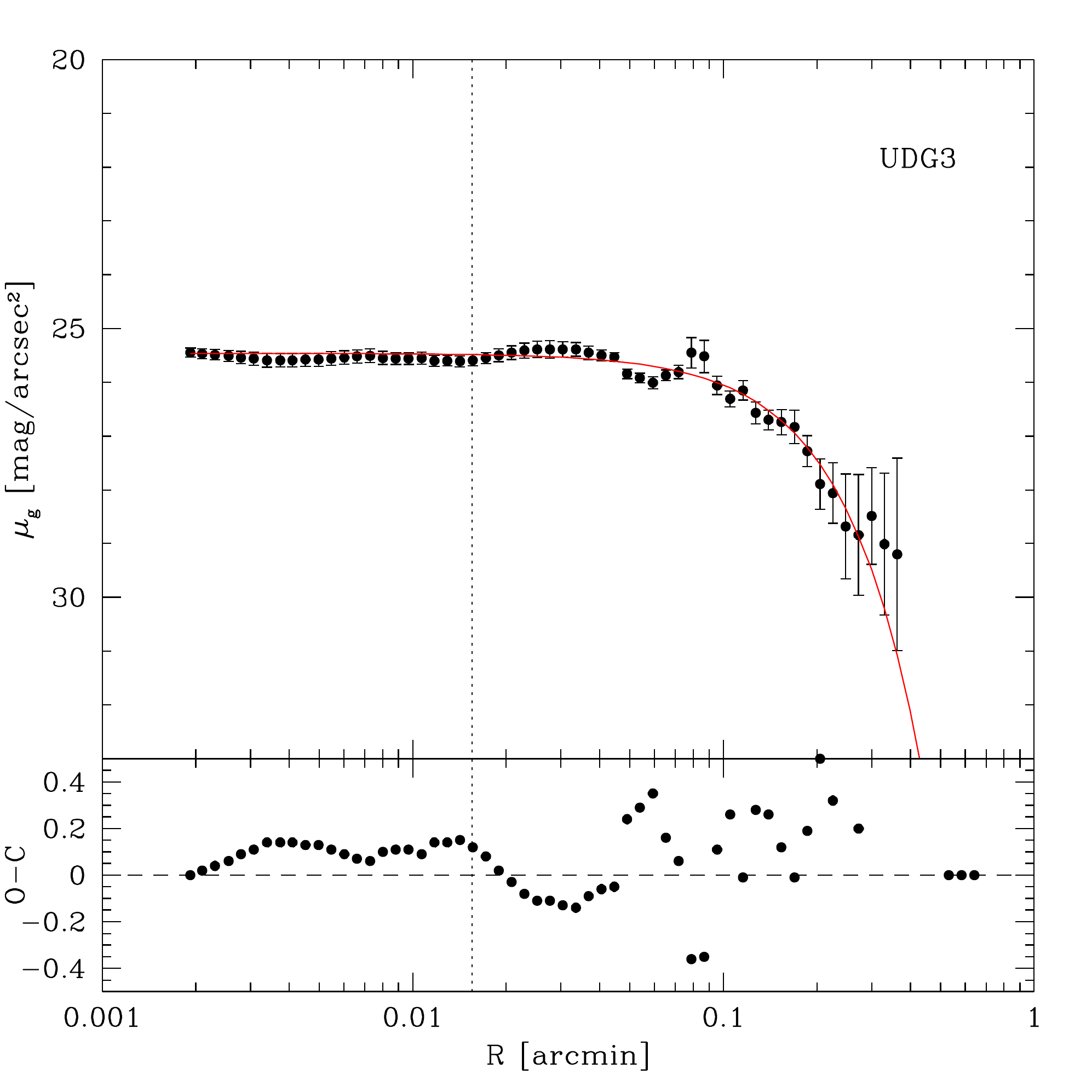}
	\includegraphics[width=6cm]{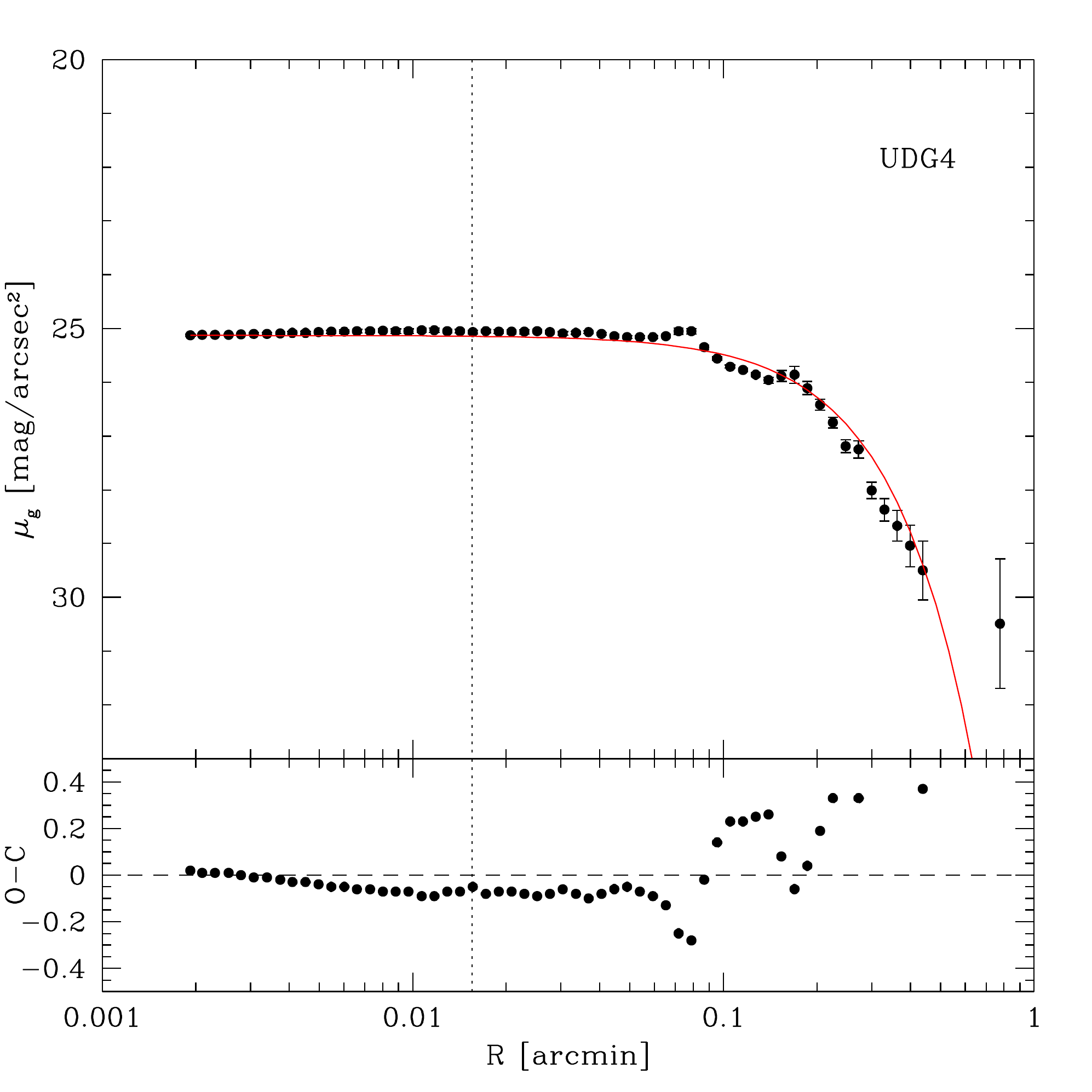}
	\includegraphics[width=6cm]{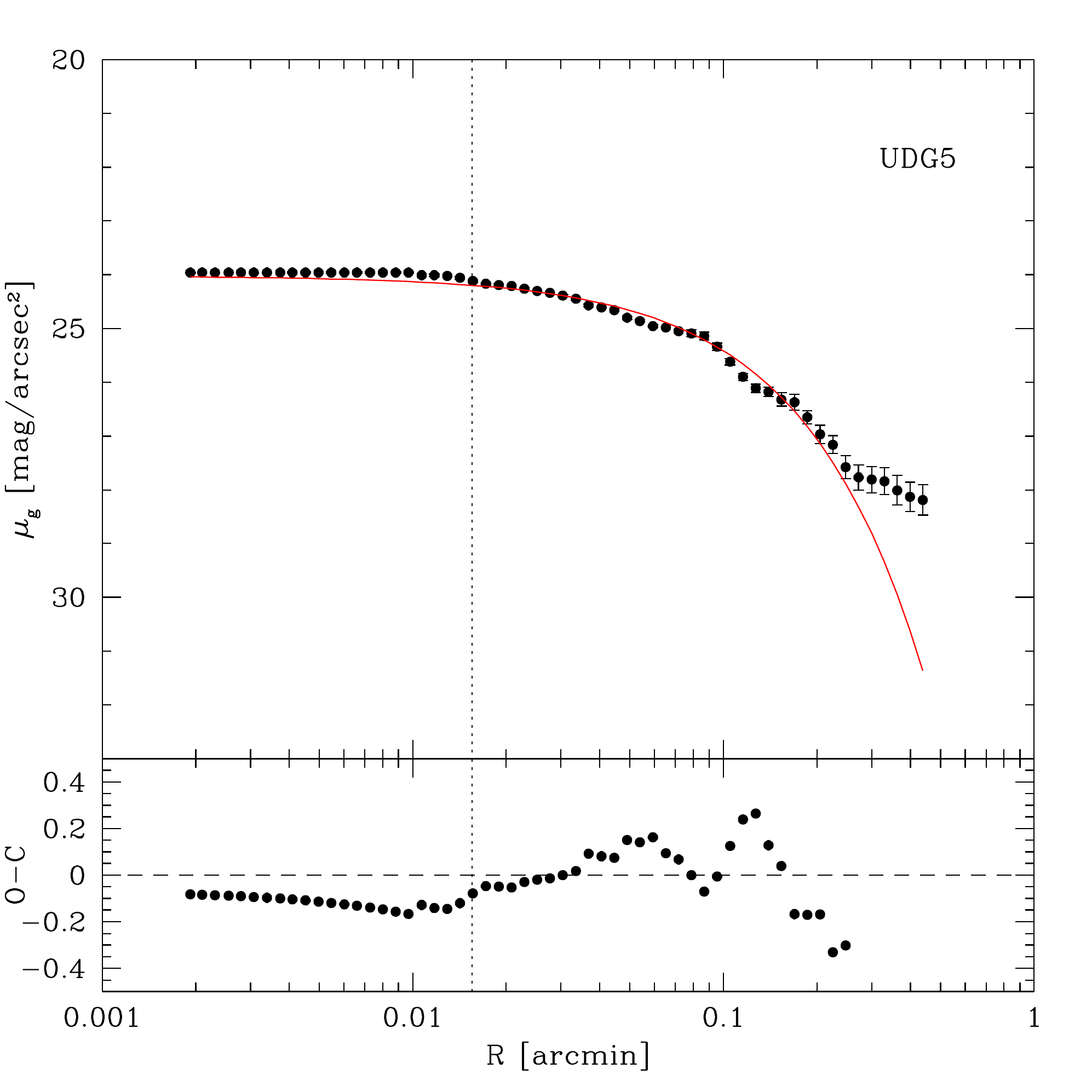}
	\includegraphics[width=6cm]{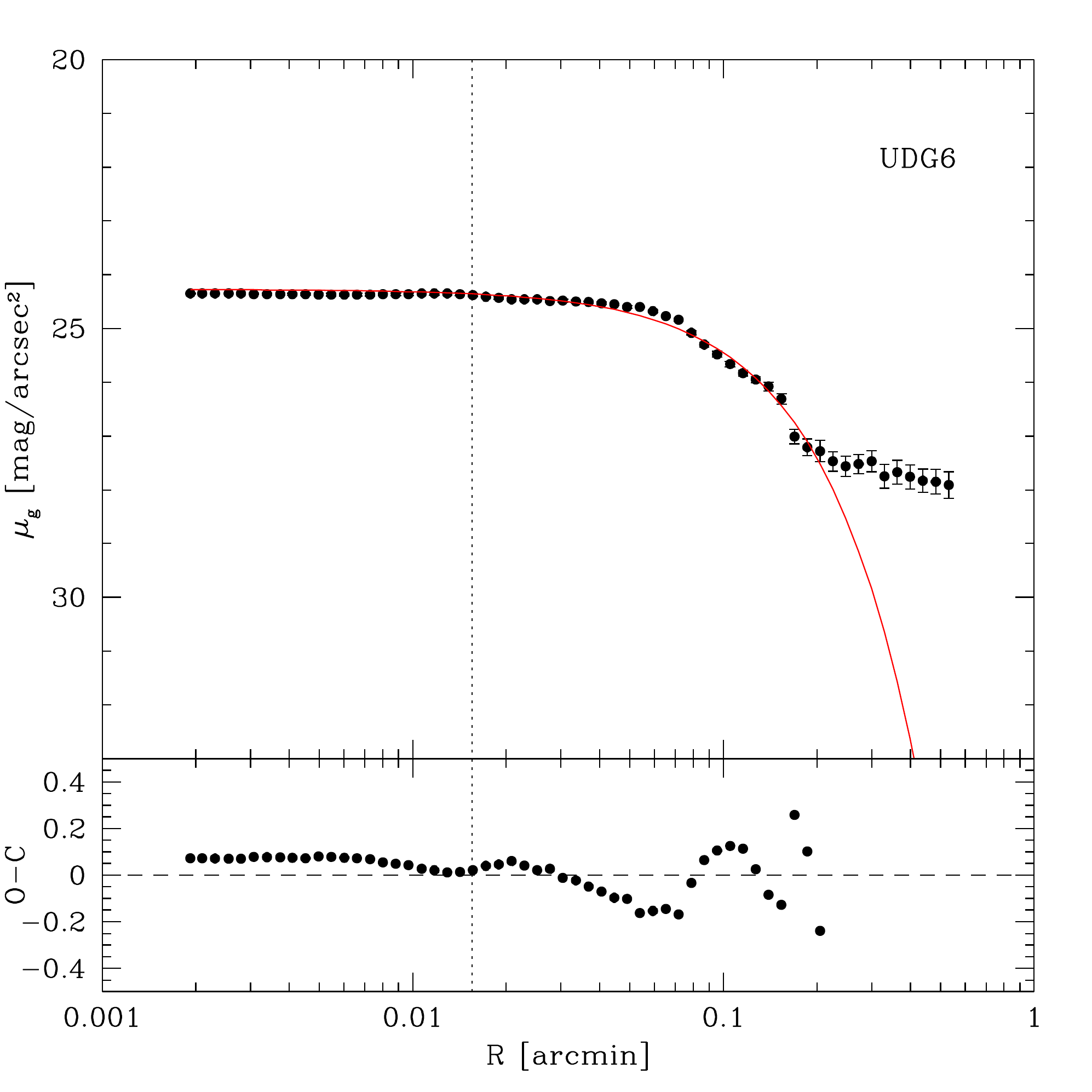}
	\includegraphics[width=6cm]{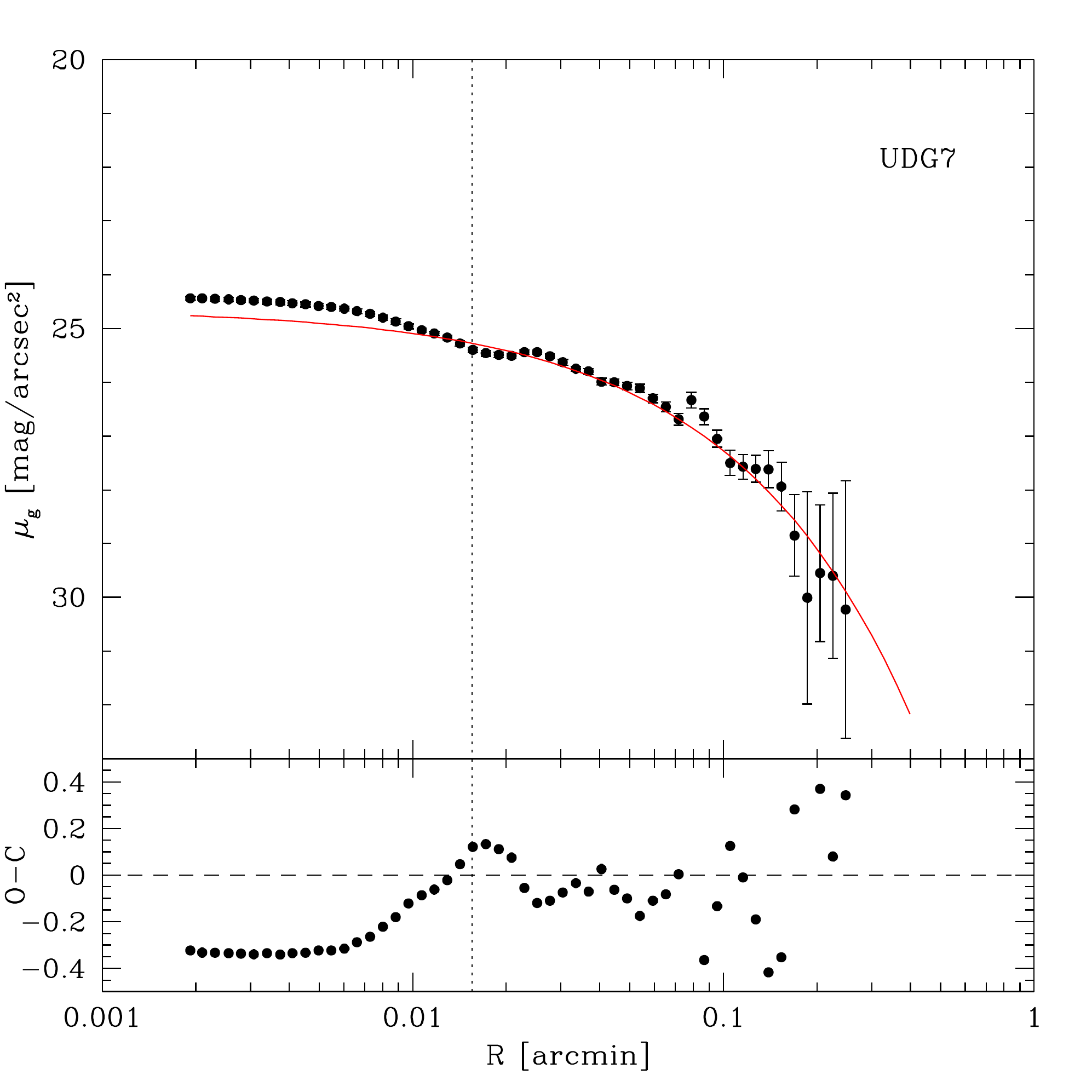}
	\includegraphics[width=6cm]{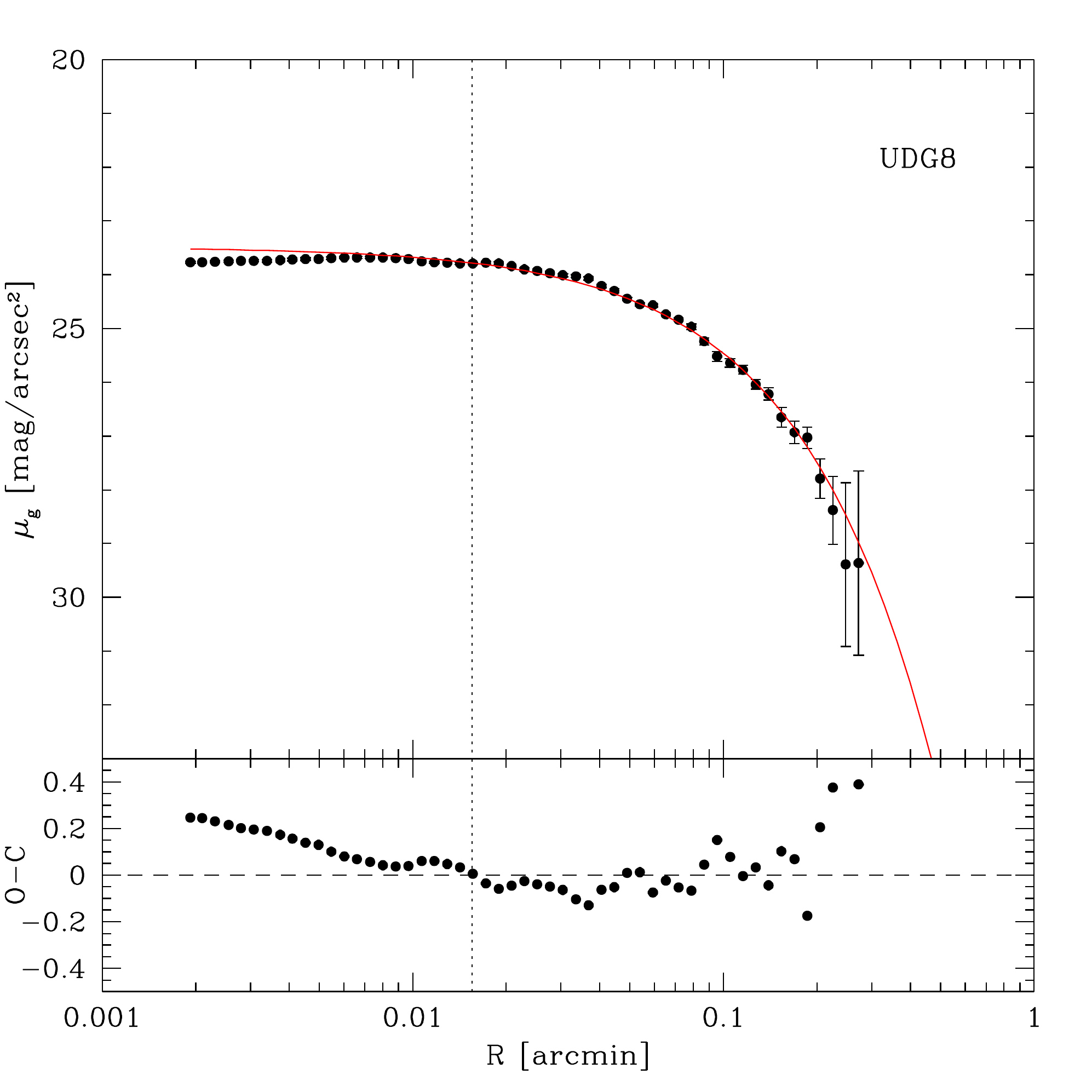}
	\includegraphics[width=6cm]{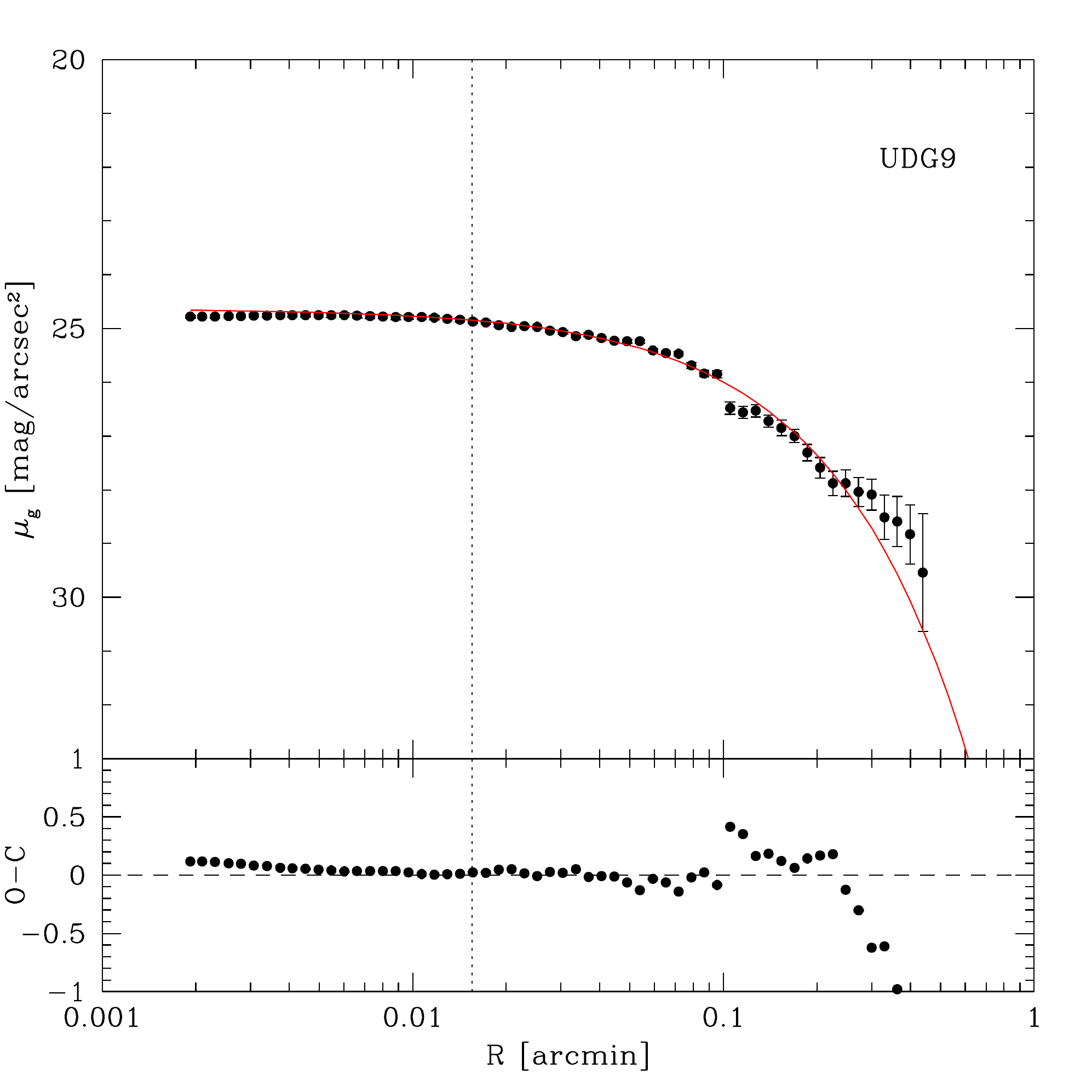}
	\includegraphics[width=6cm]{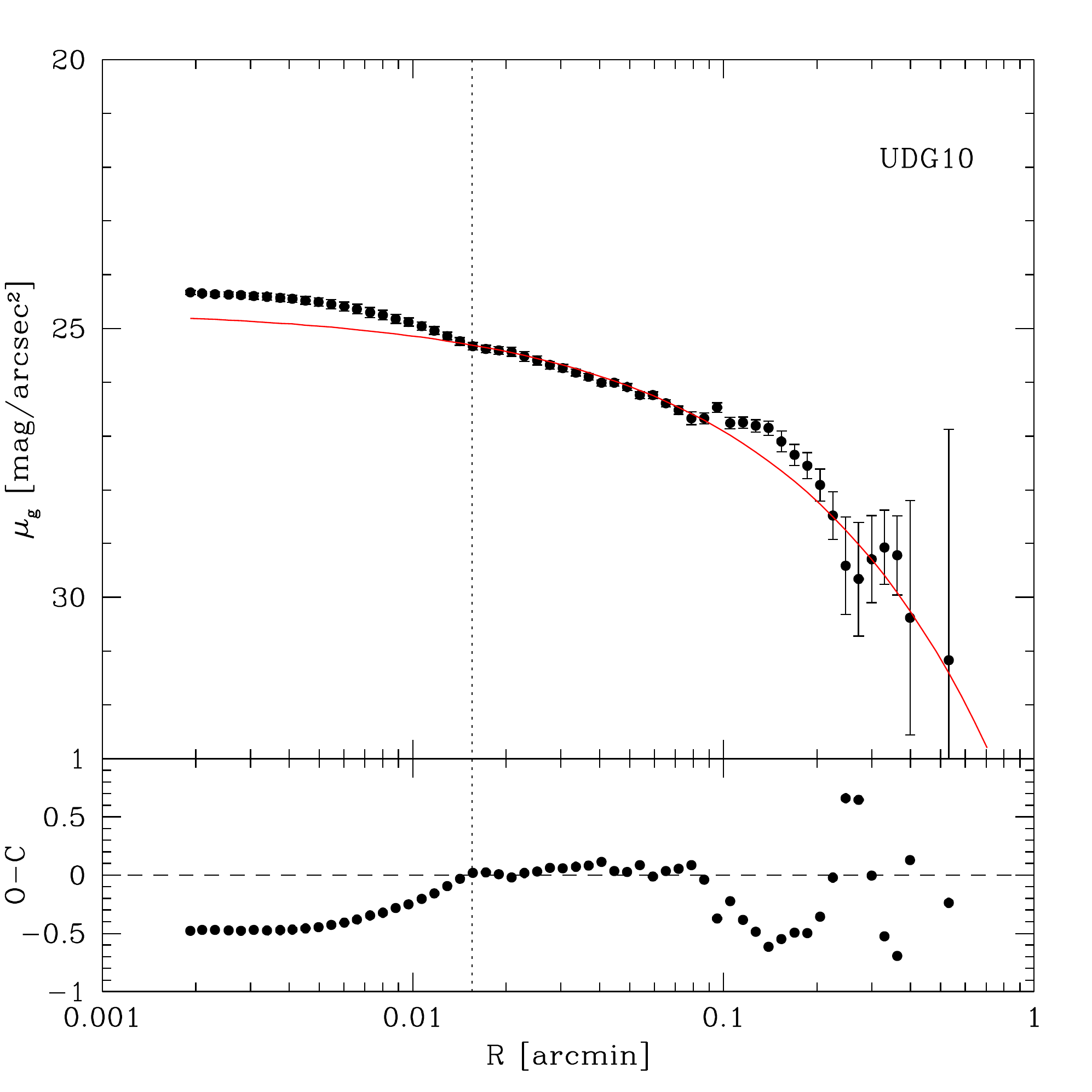}
	\includegraphics[width=6cm]{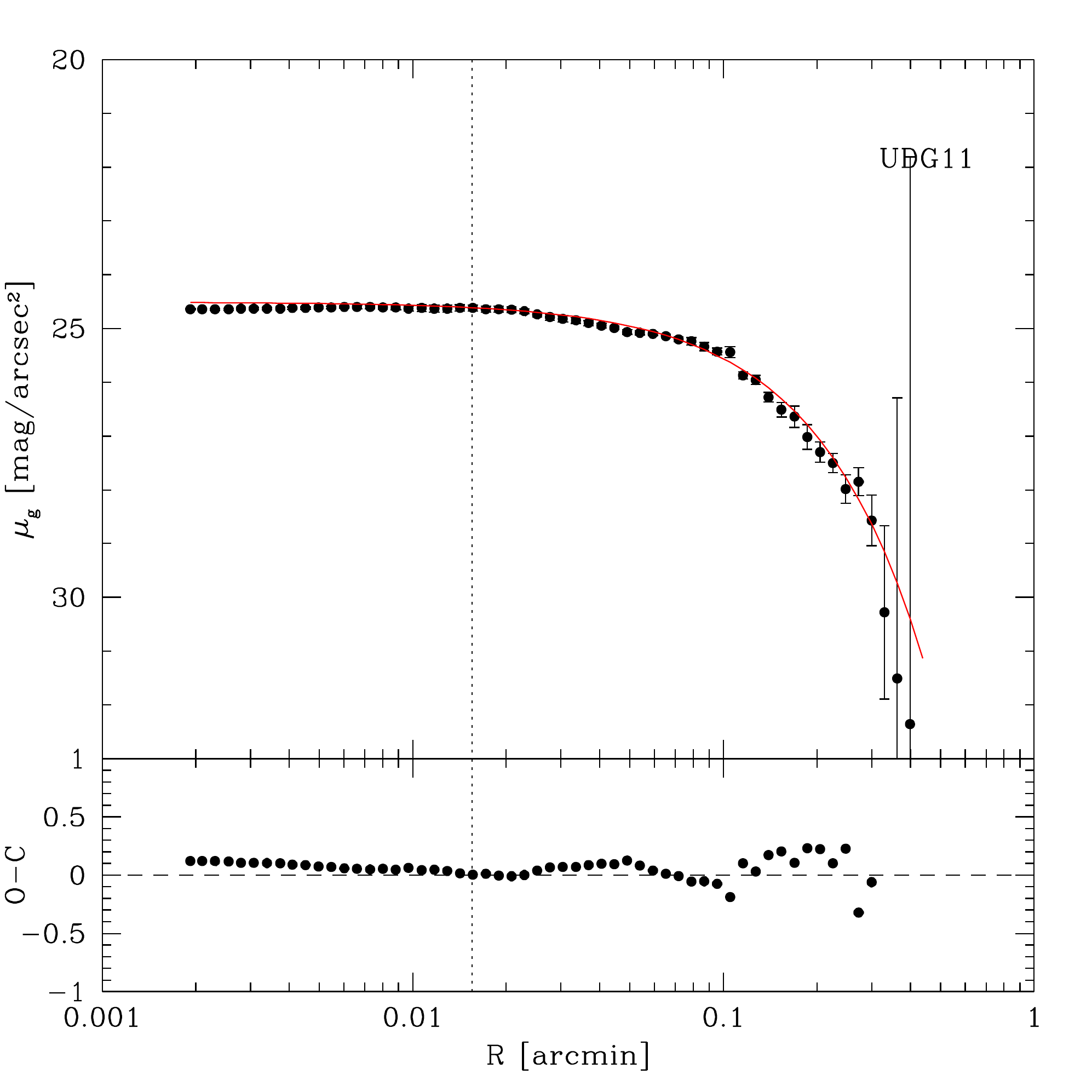}
	\includegraphics[width=6cm]{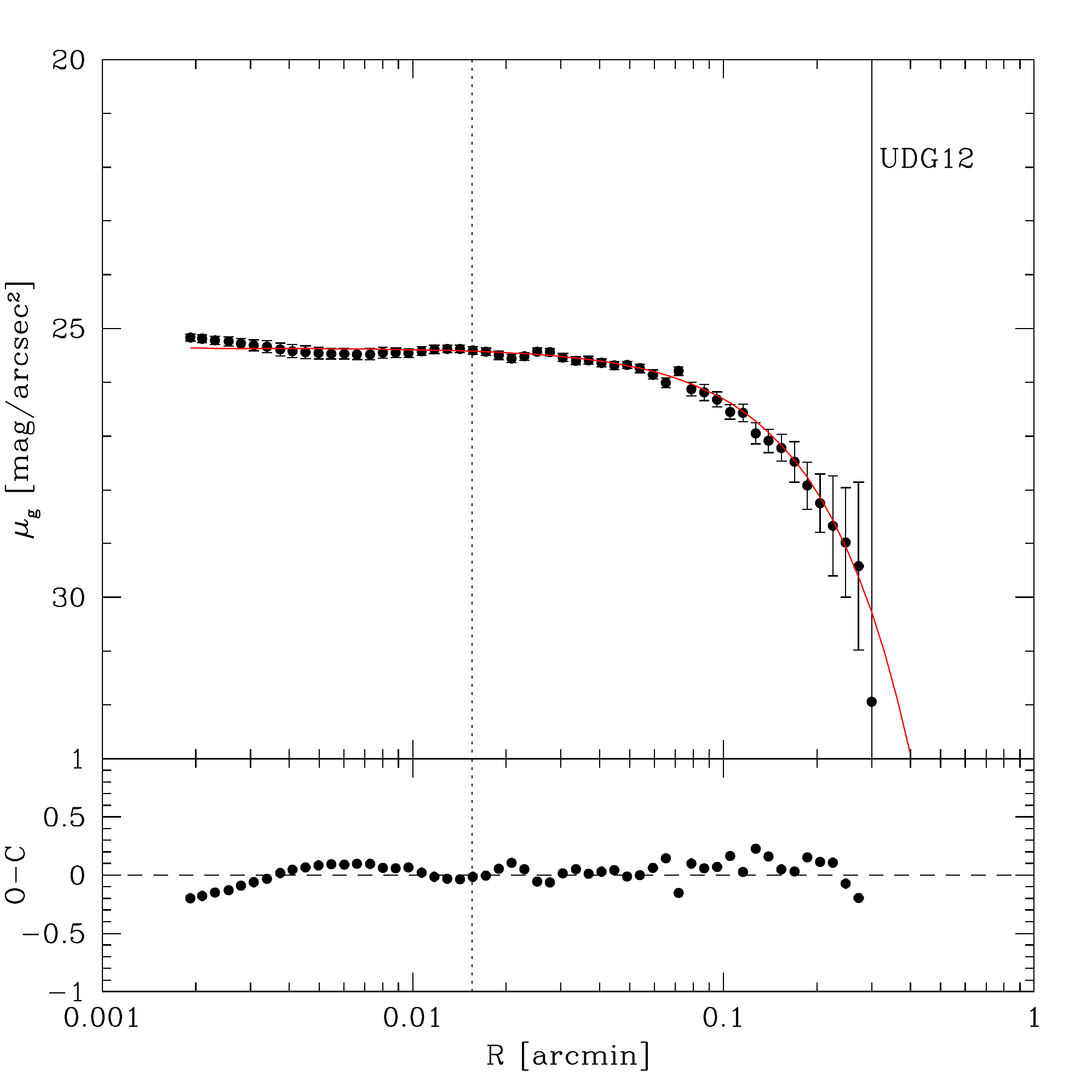}
	\caption{Azimuthally averaged surface brightness profiles in the $g$ band for the 
	UDGs candidates (top panels). The red line indicates the best-fit of the azimuthally averaged surface brightness profile derived from the 2D fit of the light distribution, using  GALFIT. The vertical dotted line indicates the inner regions excluded from the fit, inside the seeing disk. The residuals of the fit are shown in the lower panels.}
	\label{fig:UDG_prof}
\end{figure*}

Assuming a distance of $51\pm6$~Mpc  \citep{Christlein2003}, 
we adopted  the \citet{vanDokkum2015} selection criteria (i.e. $R_e\geq 1.5$~kpc and $\mu_0\geq24$~mag/arcsec$^2$ in $g$ band) 
to identify UDG candidates from the list of 27 LSB galaxies. Taking into account 
the error estimates\footnote{The error estimates on $\mu_0$ and $R_e$ take into account the 
uncertainties on the fitting, which are about 0.2\% and 6\%, respectively, 
and on the sky removal for 
$\mu_0$ ($\sim 1 \%$) and on the distance ($\sim 12 \%$) for $R_e$.} 
for $\mu_0$ and $R_e$, {  we found nine objects falling in the $\mu_0-R_e$ region 
consistent with UDGs, and additional three galaxies that cross the lower limit for the $R_e$ value, 
considering the uncertainty on the distance ($\Delta R_e \simeq 0.12$~kpc). To be as complete as possible, we also include these objects since they are consistent with the selection criteria within the uncertainties. Therefore, the final list is made of 12  UDG candidates and is given in Tab.~\ref{tab:UDGsample}.} 
All remaining LSB galaxies are listed in Tab.~\ref{tab:LSBsample}.
In both tables we include the total luminosity, colours and structural parameters.
The location of the selected UDGs inside the cluster is displayed in Fig.~\ref{fig:mosaic} with red boxes,
and the cutout of each UDG is shown in Figs.~\ref{fig:UDG_image1} and \ref{fig:UDG_image2}.
The azimuthally averaged surface brightness profiles for the UDG candidates and the relative best fits are shown in Fig.~\ref{fig:UDG_prof}. 

\subsection{Detection of the globular clusters in UDGs}\label{sec:GCs}

To identify  GCs around the UDGs we applied the same strategy already used in other works based on VST data, 
where a combination 
of the detected sources' magnitude, colours, and shape parameters is adopted to identify GC candidates \citep{cantiello2015,cantiello2018,Cantiello2020}.
%
Briefly, we proceeded as follows. We first ran SExtractor \citep{bertin96} on image cutouts centred on UDG, 
{  with size $\sim10\times R_e$ on each side}. 
To improve source detection and deblending down to the faintest magnitude level, the galaxy model derived from GALFIT was subtracted from both the $g$ and $r$ frames. For each source we derived the automated aperture magnitude based on \citet{kron80} first moment algorithms (SExtractor $MAG\_AUTO_X$, where $X$ refers to the $g$- or the $r$-band), and the aperture magnitude within 4 and 6 pixels diameters ($MAG\_APER_X$). We used $MAG\_AUTO_X$ to estimate the total magnitude of the source, while the aperture magnitudes is adopted for the $g-r$ colour $MAG\_APER_g(6pixel){-}MAG\_APER_r(6pixel)$ of the source, and the concentration index ($CI_X=MAG\_APER_X(4pixel){-}MAG\_APER_X(6pixel)$), which is an indicator of source compactness \citep{peng11}.

At the adopted distance of Hydra I, {  the turn-over magnitude (TOM) of the GC luminosity function (GCLF) is $\mu_{g,TOM}\sim26.0$ mag \citep[e.g.][]{villegas10}}, and  $\mu_{r,TOM}\sim25.4$ mag, based on a median $\langle g{-}r\rangle\sim 0.6$ (mag) for the GC population \citep{Cantiello2018a}. The TOM in both bands roughly matches with the image limiting magnitude, defined as the $5\sigma$ AB magnitude, determined from the median S/N estimated as $\Delta MAG\_AUTO_X^{-1}$.

To identify GC candidates, we selected sources with the following characteristics:
{\it i)} $g$ band magnitude between {  23.5 and 26 mag, i.e. the expected range between the TOM and 3$\sigma_{GCLF}$ mag brighter \citep{villegas10};
{\it ii)} colour within the interval  {  $0.25\leq g{-}r \leq 1.25$~mag}; 
{\it iii)} SExtractor {  $CLASS\_STAR\geq0.4$};
{\it v)} elongation, i.e. major-to-minor axis ratio $\leq2$ in both bands; 
{\it iv)} concentration index within {  $\sim0.1$} mag of the sequence of local point sources \citep[see][for more details]{Cantiello2020}.}

{  In order to select the GC members of each UDG we need to correct for the contamination of the 
sample due to foreground stars, background compact galaxies and possible intra-cluster GCs. 
This is most effectively achieved by estimating the local contamination 
in the spatial regions between $5\leq R_e\leq 10$ around each UDG candidate. The number of
contaminants subtracted from the sample of GCs selected in each galaxy is on average $3.3\pm0.8$~arcmin$^{-2}$.


We obtained two different estimates of the total number of GCs 
($N_{GC}$). Adopting the approach suggested by  \citet[][]{vanDokkum2016}, we estimated 
the total population within 1.5 $R_e$. In addition, the total number of GCs within $5 R_e$, 
which is the upper limit for bound systems \citep{Kartha2014,Forbes2017,Caso2019}, is also derived. 
Since the photometry reaches roughly the TOM peak and assuming that the GCLF 
is a Gaussian also for the UDGs  \citep[e.g.][]{Hanes1977,Rejkuba2012,vanDokkum2016}, 
we derived $N_{GC}$ as twice 
the background corrected GC density over the $5 R_e$ area of the UDG, times the 
area. The only difference with the $N_{GC}$ within 1.5 $R_e$ is that it is assumed 
that only half of the GC population is within 1.5 $R_e$, hence the total population 
is estimated as four times the background corrected GC density over 1.5 $R_e$ area, 
times the area.}

{  We also estimated the GCs background level in three independent fields located outside 
the virial 
radius of the cluster, $\sim17.5\arcmin$ wide on each side, by running the same detection and selection procedures outlined above. 
The estimated GC background density obtained is $\rho_{GC~background}=1.27\pm0.10$~arcmin$^{-2}$, a value either consistent
with the background around the UDGs, or lower than that, 
indicating the presence of GCs contaminants from close GC 
systems of major galaxies 
or due to intra-cluster GCs. We verified that the level of foreground contamination by Milky Way stars in the 
direction of Hydra~I is $\sim50\%$ of the number we estimated from real data, by using both the TRILEGAL and Besancon models \citep{girardi05,robin03}.}

The uncertainty on $N_{GC}$ is derived by propagating {  a $20\%$ error on the adopted scaling factors 
(i.e., two for $N_{GC}$ at  5 $R_e$, and four at $1.5 R_e$, respectively), a 10\% for the error on the density 
of background sources (slightly larger than the $rms$ from the three background fields outside the cluster core), 
and a $20\%$ contamination on the density of GC candidates over the UDG area. 
The two $N_{GC}$ estimates agree within the adopted uncertainties and they are listed in Tab.~\ref{tab:GCs}.
We found that six UDGs have $N_{GC}$ consistent with zero, independently from the approach adopted 
to estimate it.  Two out of the 12 galaxies (UDG~3 and UDG~11) have $N_{GC}$ different from zero at 
$> 2 \sigma$ level. 
One object (UDG~4) has $N_{GC}$ different from zero at only $1 \sigma$ for both estimates. 
All other UDGs in the sample have $N_{GC}$ between 1 and 2 $\sigma$ different than zero, 
depending on the approach used (UDG~2, UDG~7, UDG~9).

$N_{GC}$ is used to estimate {\it i)} the total halo mass $M_h$ for those UDGs with a significant number of GCs,} by using the
empirical relation $log[M_h] = 9.68+1.01\times log[N_{GC}]$ \citep{Burkert2020}, and {\it ii)} the GCs specific frequency\footnote{  The V-magnitude and, therefore, the colour transformation from $g-r$ to V-I are derived by assuming the equations given by \citet[][]{Kostov2018}.} 
$S_N = N_{GC} 10^{0.4 [M_v+15]}$ . 
Values for both quantities are given in Tab.~\ref{tab:GCs}.
{  The level of uncertainty on $N_{GC}$ and the number of GC systems consistent with $N_{GC}=0$ 
for half of UDGs in the sample imply that our estimated $M_h$ and $S_N$ suffer from correspondingly 
large uncertainties, and are in most cases formally consistent with zero. 
In future works, with more detailed characterisation of the GCs population extended over the survey area, 
we will be able to perform a more refined analysis of the GC numbers.}


\section{Results: UDGs structure and mass content}\label{sec:results}

\begin{figure*}
	\centering
	\vspace{-7cm}
	\includegraphics[width=\hsize]{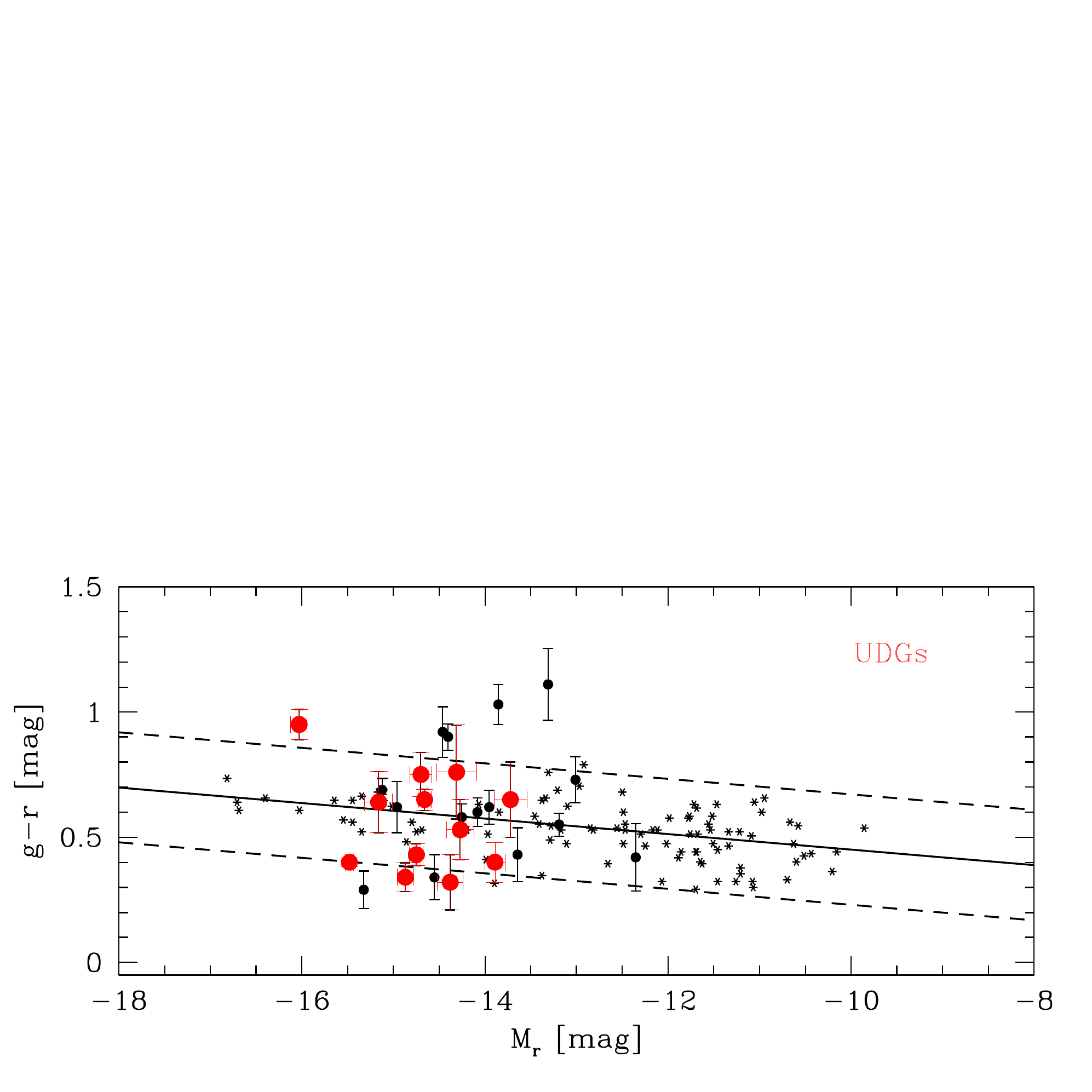}
	\caption{colour-magnitude diagram (CMD) for the full sample of LSB galaxies (black points) detected in the VST Hydra I mosaic. 
	Red filled circles indicate the UDG candidates. The solid black line is the CM relation for the Hydra I cluster galaxies derived by \citet[][dashed lines indicate the {  2$\sigma$} scatter]{Misgeld2008}.  The asterisks indicate  dwarf galaxies from \citeauthor{Misgeld2008}. }
	\label{fig:CMD}
\end{figure*}

The colour-magnitude diagram (CMD) of the full sample of LSB galaxies and 
UDGs visually identified is shown in Figure 
\ref{fig:CMD}. The figure also includes the colour-magnitude relations derived by \citet{Misgeld2008}  for giant 
and dwarf galaxies in Hydra I. In this plane, based on the error estimates, 25 out of 27 LSB galaxies match with the plotted sequence.

{  In the range of magnitudes $M_r \simeq -15.5$~mag to $M_r \simeq-13.5$~mag, the $g-r$ colours of UDGs are consistent with the known early-type dwarf galaxy population in Hydra I, with $0.3 \leq g-r \leq 0.8$~mag.
Similar colours, in the same range of luminosity, were found for the dwarf galaxies in the Fornax cluster \citep{Venhola2019}, as well as for 
the UDGs in the Abell~186 cluster \citep{Roman2017a}, i.e. $0.5 \leq g-r \leq 0.8$~mag.} 


{  UDG~4, the reddest and brightest LSB galaxy in the sample 
($g-r=0.95\pm0.10$~mag, $M_r=-16.04$~mag, Tab.~\ref{tab:UDGsample}), 
being $\sim 0.1$ mag 
beyond the $2\sigma$ boundary of the CMD, might rather be a background galaxy. 
Unfortunately, the distance for this galaxy is unknown. 
However, even though we cannot derive  any definitive conclusion on
its membership, since this is one of the most extended and diffuse galaxies of the whole 
LSB sample (with $R_e=2.64$~kpc, $\mu_0=24.86$~mag/arcsec$^2$, see Tab.~\ref{tab:UDGsample}), it
remains an interesting object to include in our analysis.}

\begin{figure*}
	\centering
	\includegraphics[width=\hsize]{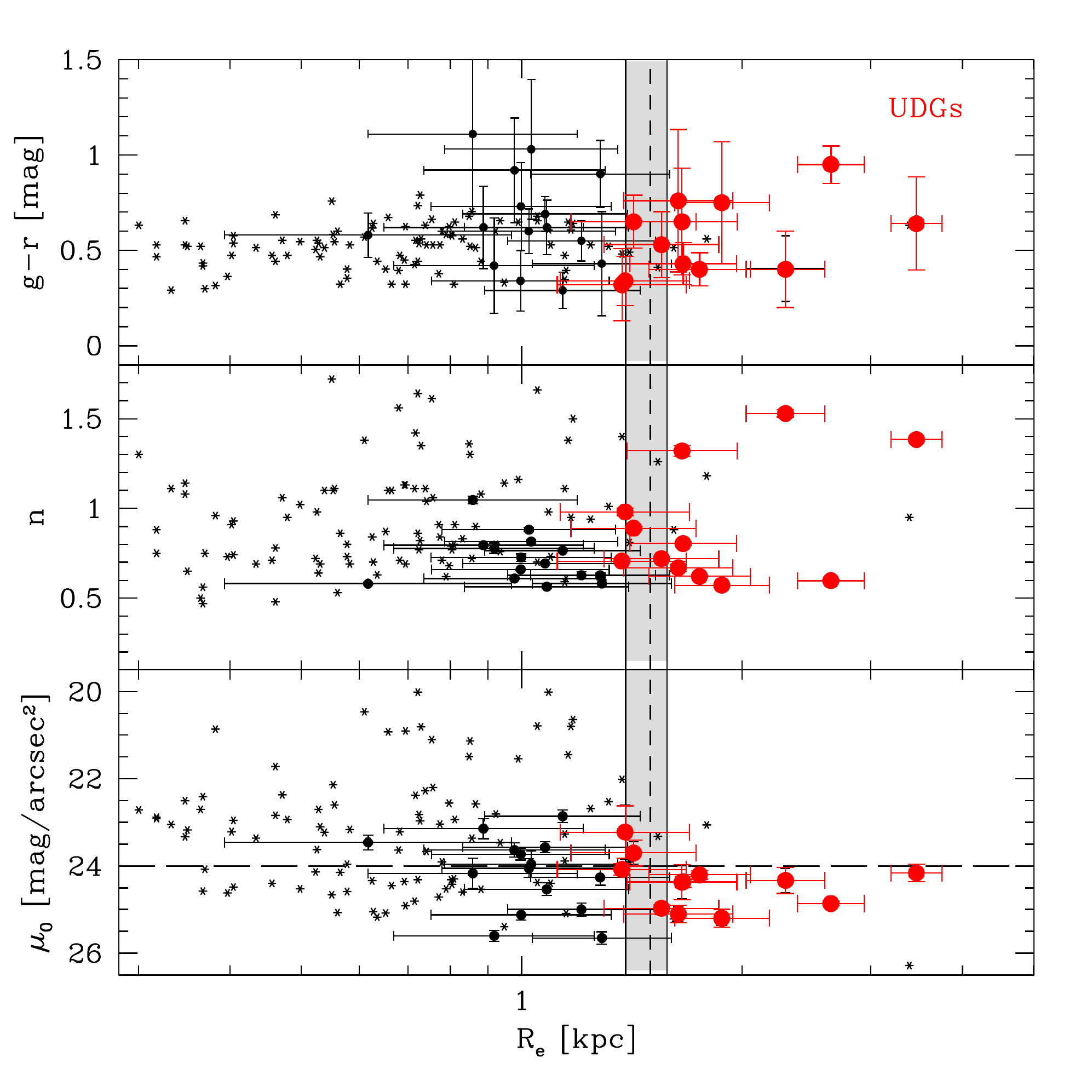}
	\caption{Structural and photometric parameters for the newly discovered LSB galaxies (dots with error-bars) in  Hydra I as a function of the effective radius. 
	The UDGs are marked with red circles. The UDG definition criteria, $R_e\geq1.5$~kpc and $\mu_0\geq24$~mag/arcsec$^2$ \citep{vanDokkum2015}, are shown by the dashed lines. 
	{  The shaded box indicates the lower and upper limits on the selection criteria on $R_e$ due to the uncertainty on the distance (i.e., $R_e \pm 0.12$~kpc). }
	The asterisks are  dwarf galaxies in  Hydra I from \citet[][]{Misgeld2008}. 
	{  The only LSB galaxy in  \citet[][]{Misgeld2008} catalogue falling in the region of UDGs is HCC~087. As explained in the text, this  was described as a faint tidally disrupting dwarf 
	because of its peculiar S-shape \citep{Koch2012}.}}
	\label{fig:corrRe}
\end{figure*}

The detected LSB candidates and the selected UDGs are plotted on the $\mu_0 - R_e$ plane, in 
Fig.~\ref{fig:corrRe}. In this figure we also plot the dwarf galaxies from the \citet[][]{Misgeld2008} catalogue. 
All our LSB galaxies are new candidates, fainter and more extended, with respect to the objects in that catalogue.  
%
%
Only one LSB galaxy in  \citet[][]{Misgeld2008} catalogue, HCC~087, falls in the selection region of UDGs. However, it 
 was described as a faint ($\mu_0\sim26.2$~mag/arcsec$^2$ in the $g$ band) tidally disrupted dwarf because of its peculiar S-shape \citep{Koch2012}.
 
In Fig.~\ref{fig:corrRe} (middle and upper panels) we examine the correlations 
between the Sersic $n$-exponent and average $g-r$ colours as a function of $R_e$, 
for both the new LSB 
galaxies (including UDGs) and the dwarf galaxies from \citet[][]{Misgeld2008}. 
Most of the LSB galaxies and UDGs have comparable colours with those observed for cluster dwarfs, 
in the range $g-r\sim0.3-0.8$~mag. The Sersic $n$-exponents derived for galaxies in the sample presented here
are consistent with the values for dwarfs ($n\sim0.4-1.8$).

\begin{figure*}
	\centering
	\includegraphics[width=\hsize]{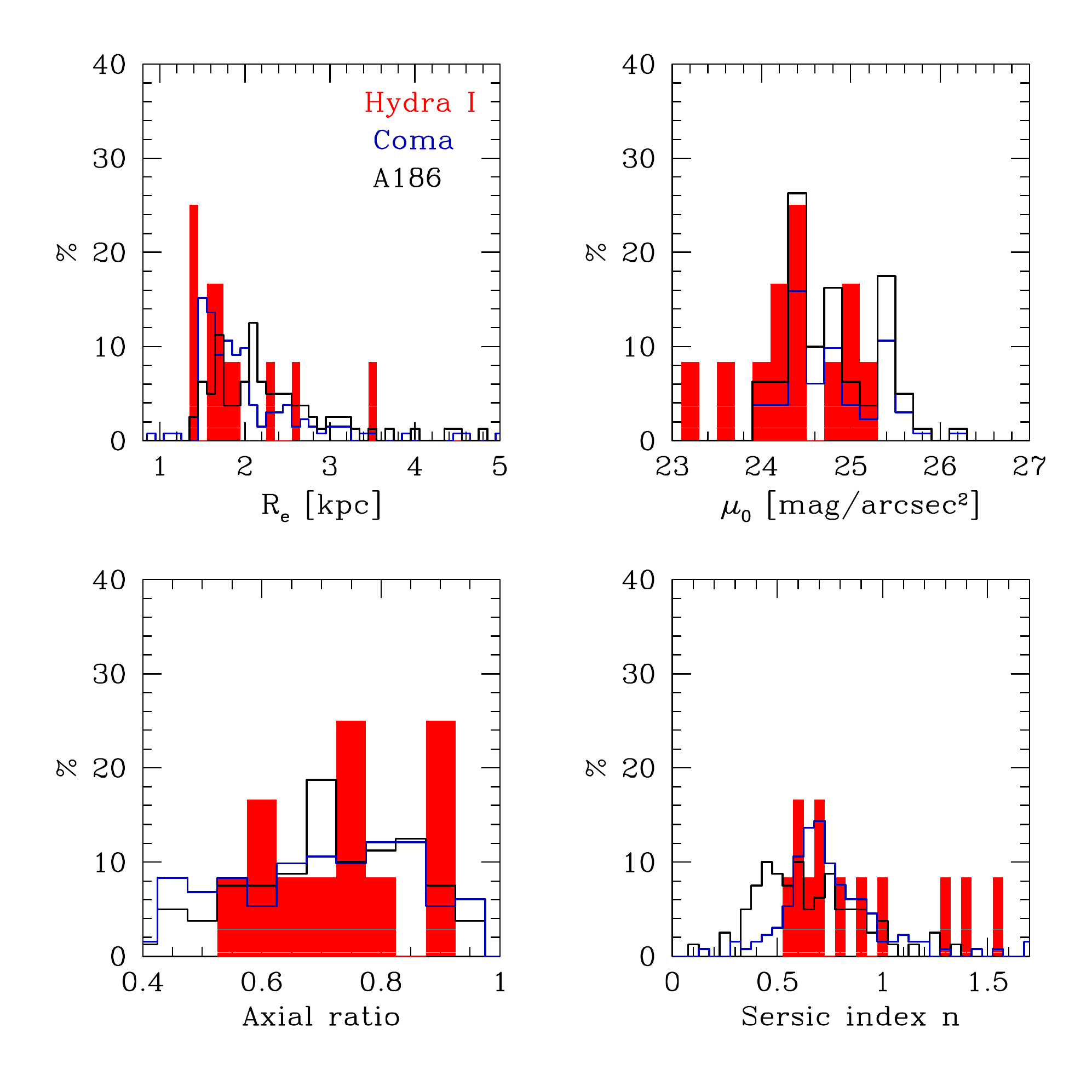}
	\caption{Distribution of the structural parameters for the  UDG candidates in Hydra I  ({  filled red histogram}). Axial ratio (lower-left panel), Sersic index $n$ (lower-right panel), effective radius $R_e$ (top-left panel) and central surface brightness $\mu_0$ (top-right panel) are compared with the same parameters derived for the UDGs in the Abell~168 cluster (black line) by \citet{Roman2017a} and in Coma cluster (blue line) by \citet{Yagi2016}.}
	\label{fig:conf_hist}
\end{figure*}

In Fig.~\ref{fig:conf_hist} we show the distributions of the structural parameters (axial ratio $q$, Sersic index $n$, 
effective radius $R_e$ and central surface brightness $\mu_0$) derived for the UDG candidates in Hydra~I cluster.
They are consistent with those 
obtained for the UDGs in the Coma and Abell~168 clusters \citep{Yagi2016,Roman2017a}, 
as well as with the results from more recent studies on UDGs in clusters of galaxies \citep{Lee2020}. 
As observed in other clusters, UDGs in Hydra~I seem to be quite round systems, 
with $q$ in the range 0.6-0.8, and most of them have a Sersic index $n \sim 0.5-1.0$, with an average value $n\sim0.8$, and only a few UDGs have larger values for the $n$-exponent.

\subsection{ Structure and colour distribution } 
\begin{figure*}
	\centering
	\includegraphics[width=\hsize]{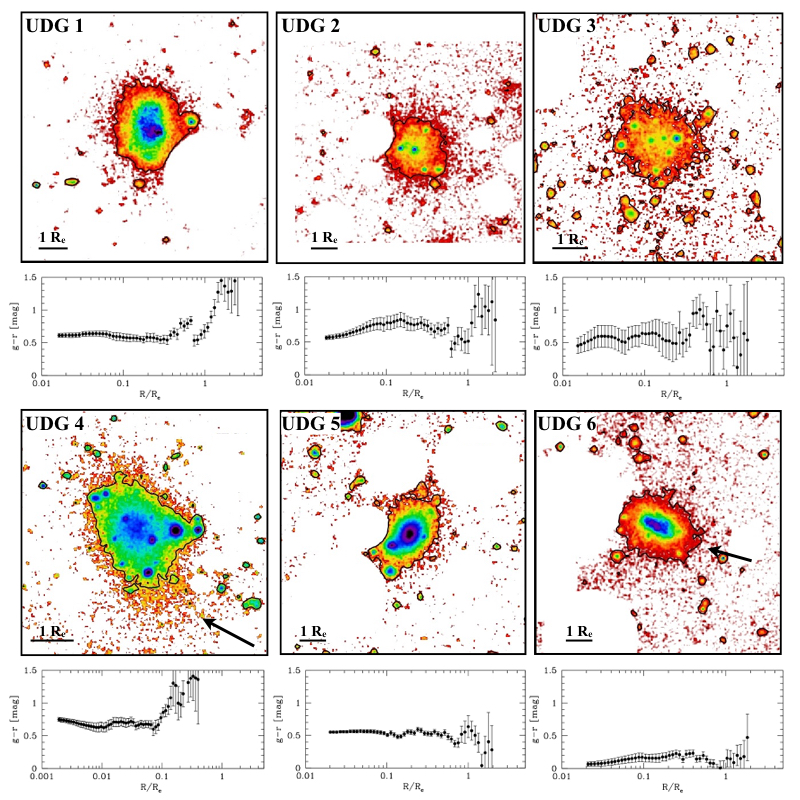}
		\caption{Images of the UDG candidates in the $g$ band in flux levels. The black contour indicates the surface 
	brightness level of $\mu_g = 27$~mag/arcsec$^2$.  In the lower-left corner we show the $R_e$ lenght. 
	The field of view of all boxes is $1\times 1$ arcmin ($\sim 14.9$ kpc). The black arrow in the image of UDG~4 and UDG~6 indicates the tidal features described in the text. The 
	area that was masked due to contaminating stars is shown as white (circular) background. 
	Below each image we plot the $g-r$ 	colour profile of the relative UDG.}
	\label{fig:UDG_image1}
\end{figure*}

\begin{figure*}
	\centering
	\includegraphics[width=\hsize]{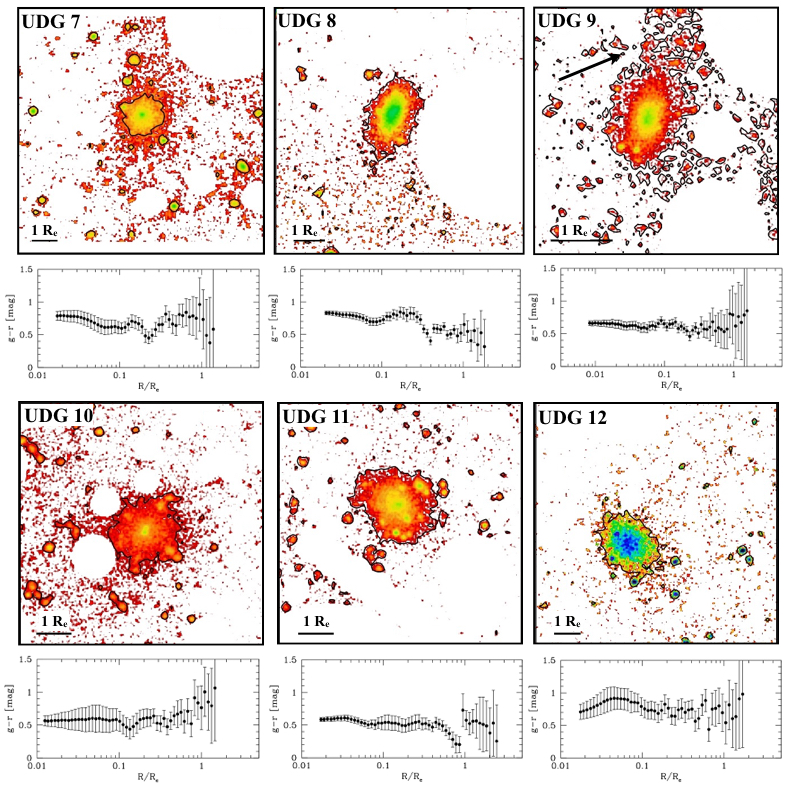}
		\caption{ Same as Fig.~\ref{fig:UDG_image1}.
 The black arrow in the image of UDG~9 indicates the tidal tail described in the text.}
	\label{fig:UDG_image2}
\end{figure*}

The deep VST images allow us to map the surface brightness distribution of the UDGs down 
to $\mu_g \sim 27-29$ mag/arcsec$^2$ (see Figs.~\ref{fig:UDG_image1} and \ref{fig:UDG_image2}). 
At these depths, we are able to study the galaxy outskirts and detect any signs of tidal features. 

This seems to be the case of UDG~4, UDG~6 and UDG~9.
The structure of UDG~4 is quite irregular in the outskirts, showing a clear over-density of light on the SW side. 
UDG~6 and UDG~9 have more regular spheroidal-like shapes, however they present tidal features in the outskirts 
and are located towards the central part of the Hydra I cluster, as shown by the brighter X-ray emission \citep{Hayakawa2004,Hayakawa2006}. 
We describe their outer  morphologies hereafter:  
UDG~6 has spiral-like tails on the SW side. It is located North of the cluster core (see Fig.~\ref{fig:mosaic}), where tidal forces might be acting on the galaxies: on the west there is HCC~005, an S0 galaxy with a prominent tidal tail, and it is near the tidally disrupted dwarf galaxy HCC~087 \citep{Koch2012}.
UDG~9 has an extended tail ($\sim0.5$~arcmin) on the NW side that protrudes from the 
elongated isophotes of this galaxy in this direction. 
UDG~9 is projected on the SE region of the cluster core,
inside the infalling subgroup dominated by the bright spiral galaxy NGC~3312, 
where the effect of ram pressure is visible in the form of several extended blue filaments. 
Tidal features around UDGs have also been observed in other clusters for objects that are close to  
major galaxies \citep{Mihos2015a,Lee2017}. 
Such features would support the tidal interaction formation scenario {  \citep{Amorisco2018,vanDokkum2016,diCintio2017,Carleton2019,Sales2020}}.

Close to the centres of UDG~1 and UDG~5 the are signs of an ongoing interaction. 
In UDG~1 we detected a spiral-like feature that ends with a bright nucleus on the west.  
In UDG~5 there is a small system on the SE side that seems to be merging into the galaxy core. 

The  $g-r$ colour profiles are on average flat inside $1 R_e$ for most of the UDGs, in the range 
$0.3-0.7$~mag (see Fig.~\ref{fig:UDG_image1} and Fig.~\ref{fig:UDG_image2}). 
UDG~4, UDG~7, and UDG~8 show redder colours toward the centre, reaching $g-r\sim0.8$~mag. 
Some other UDGs (UDG~2, UDG~3, UDG~6 and UDG~12) have instead bluer colours in the centre, 
decreasing by $\sim0.2$ mag with respect to the outer regions. 


\subsection{ Compact sources in UDGs} 
For most of the UDGs we were able to to identify compact sources in their surroundings.
(see Sec.~\ref{sec:GCs}). 
{  Overall, we found that most of the UDGs host only a few GCs, with $3\leq N_{GC} \leq 10$
for six out of 12 UDG candidates inside $5 R_e$ (see Tab.~\ref{tab:GCs}). 
UDG~3 and UDG~9 are the two objects  with the largest number of 
GCs, having $N_{GC}=6$ and $N_{GC}=10$ inside $5R_e$, respectively.
For six UDGs, the predicted number of GCs is consistent with zero, within both $1.5 R_e$ and $5 R_e$. }

\begin{figure*}
	\centering
	\includegraphics[width=\hsize]{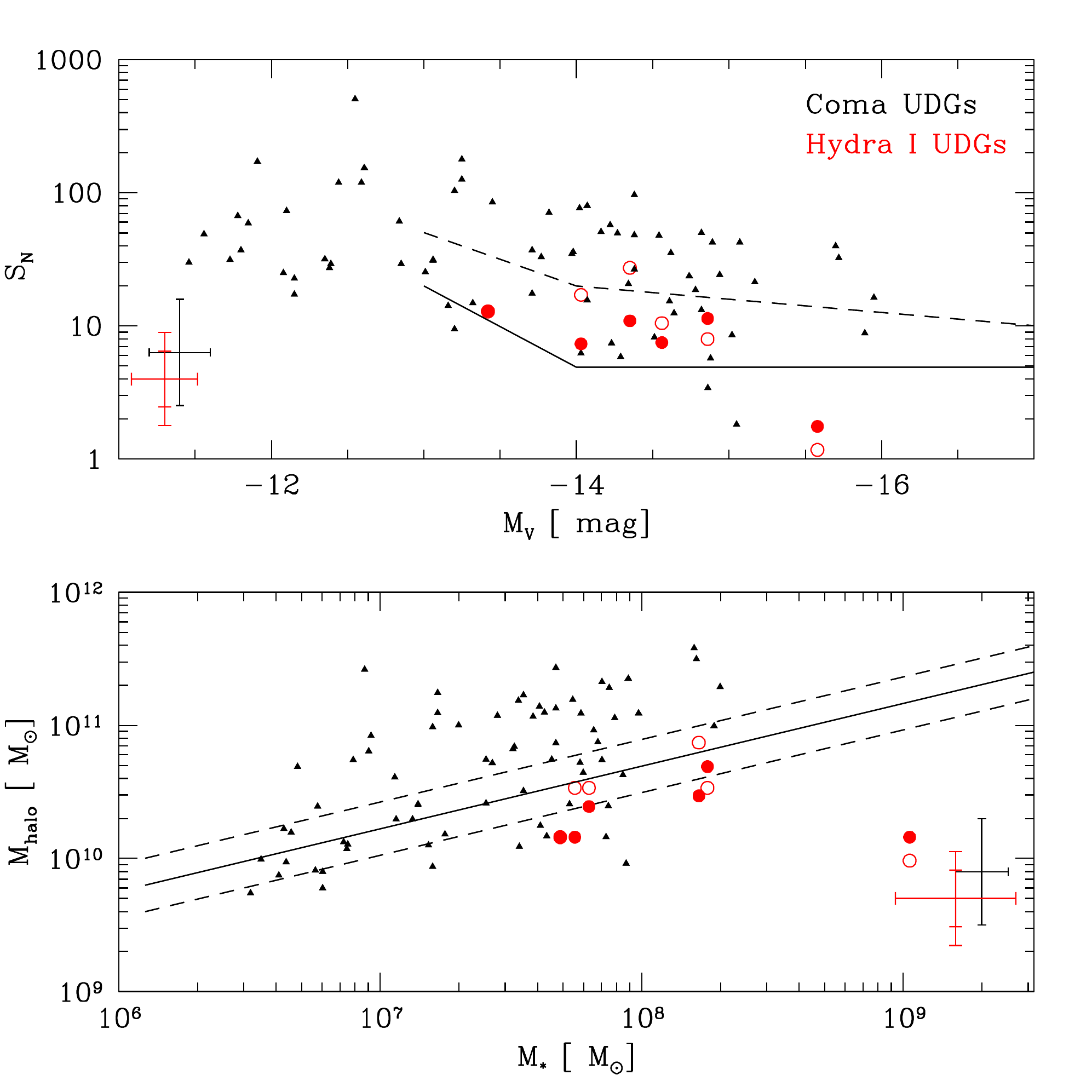}
	\caption{
	{\it Upper panel -} GCs specific frequency $S_N$ versus $V$-band absolute magnitude for the Hydra~I UDG sample (red), compared with the UDGs in the Coma cluster (black triangles) by \citet{Forbes2020a}. For the 
	UDGs in Hydra I,  {  the filled red circles are for values derived inside $5R_e$}, empty red circles for values derived inside $1.5R_e$. The solid line shows the mean locus of  dwarf galaxies, the dotted line represents the upper 2$\sigma$ bound \citep[see also Fig.4 in][]{Lim2018}. 
	The average uncertainties on both samples are shown in the lower-left corner.
	{\it Lower panel -} Halo mass versus stellar mass. The halo for the UDGs in Hydra I is derived from the total number of GCs, using the scaling relation by \citet{Burkert2020}, see text for details. Symbols are the same as in upper panel. The average uncertainties on both samples are plotted in the lower-right corner. The black solid line is the extrapolated stellar mass-halo mass relation for normal galaxies by \citet{Rodriguez2017}, with a scatter of $\pm 0.2$~dex (dashed lines).}
	\label{fig:hmass}
\end{figure*}

As observed in previous studies with larger samples, we found that 
for most of the UDGs in Hydra I,
the GCs specific frequency $S_N$ is {  consistent with the upper limits derived for dwarf 
galaxies of comparable luminosity} 
(see top panel of Fig.~\ref{fig:hmass}), and it is fully consistent with values derived for UDGs in 
the Coma cluster {  \citep[see][]{vanDokkum2017,Amorisco2018a,Lim2018,Forbes2020a}}. 
{  UDG~3, UDG~7 and UDG~9 are the galaxies with the largest $S_N$ ($\sim 11-13$). 
Two of them, UDG~7 and UDG~9, are close in projection to the cluster core. 
While the actual $S_N$ for our targets needs to be confirmed through follow-up studies, this appears in agreement with suggestions from  \citet[][]{Lim2018}, who reported,  
although  with a large scatter, a trend between $S_N$ and environment, 
where UDGs with higher $S_N$ are located in the densest cluster regions.} 

Consistent with previous results, about 30\% of the UDGs
in our sample shows a nuclear star cluster (NSC) candidate, which 
are the brightest 'GCs' in their respective GC systems and are located very close to the galaxy centre.
As a comparison, the nucleation fraction  is about 23\% for UDGs in Coma \citep{Lim2018}.
The two NSC candidates in UDG~2 are very close to the galaxy centre ($\leq 1$ arcsec) and have a 
$m_g\simeq 24.2$ and $25.2$~mag. UDG~3 hosts a NSC candidate inside 1 arcsec from the centre, with $m_g\simeq 
24.2$~mag, though also a second, brighter $m_g\simeq 22.2$~mag, source is consistent with a star cluster
located at $\sim 7$~arcsec 
from the galaxy centre (see also Fig.~\ref{fig:UDG_image1}). The candidate NSC in UDG~7 is in the galaxy centre and has a relatively faint g-band magnitude of 
$m_g\simeq 26.2$~mag. In UDG~10 there are two close sources within the galaxy centre that could be also classified as NSCs.
%

\subsection{ Stellar mass versus DM content } 
The $g-r$ colours and absolute magnitudes in the $r$ band (M$_r$) 
are used to derive the stellar mass for all UDG candidates, using the relation 
given by \citet{Into2013}. Values are listed in Tab.~\ref{tab:UDGsample}.
Most of the UDGs have stellar masses in the range $10^{7}-10^{8}$~M$_{\odot}$. 
{  The halo mass range, estimated from the total number of GCs (see Sec.~\ref{sec:GCs}) for 
half of the UDGs in the sample, is $1-3\times 10^{10}$~M$_{\odot}$ (see Tab.~\ref{tab:GCs}). By propagating the large uncertainties on $N_{GC}$, the error
estimate on $M_h$ is quite large, ranging from 40\% to 80\%.
The mass-to-light ratio derived for these six UDGs is in the range $250 \leq M/L_V \leq 10^3$.
UDG~3, which is one of the galaxies with the highest $S_N$ (see Tab.~\ref{tab:UDGsample} and Tab.~\ref{tab:GCs}), has the largest $M/L_V \sim 10^3$. 
The lowest $M/L_V \sim34$ is derived for  UDG~4, which is also the system with the lowest 
$S_N\sim 2$. Therefore, if UDG~4 is confirmed as Hydra I member
 in future investigations, it might have a low amount of DM compared to what is expected for its total luminosity.
However, the estimated $N_{GC}$ in this case is consistent with zero within $1\sigma$, therefore, 
given its high luminosity, the absence of GCs would point toward this galaxy being a background object. 

In the halo mass versus stellar mass relation, within the uncertainties, 
the UDGs in Hydra I have halo masses consistent with 
normal galaxies of comparable luminosity, 
as well as with UDGs in the Coma cluster (see lower panel of Fig.~\ref{fig:hmass}). }

\section{Summary and Concluding remarks}
From the visual inspection of a new mosaic image obtained for the Hydra~I cluster, 
with VST $g$ and $r$ imaging, 
we identified a sample of 27 LSB galaxies, which are
not included in any previous catalogue.
12 LSB galaxies are selected as UDG candidates.
This is the first sample of UDGs in the Hydra~I cluster.

Since the nature and formation of the UDGs is still poorly constrained, 
even from works based on larger samples, 
this study aims at extending such studies to the Hydra~I cluster, which had 
so far limited studies in the  low surface brightness regime.

{  We have found that most  of  the  UDGs have  stellar  masses  in the range  $10^7-10^8$~M$_{\odot}$. 
Given the limitations of a reliable GC selection based on two relatively close optical bands only,  
we find that half the UDG candidates have a total population of GCs consistent with zero ($N_{GCs}\sim0$). 
The other half of UDGs, seem to have a standard or low DM content, with halo mass $\leq 10^{10}$~M$_{\odot}$, comparable with dwarf galaxies of similar luminosity. }


{  In conclusion, even considering the small number of UDGs, the analysis presented in this 
work might suggest that most of the UDGs in Hydra~I cluster resemble diffuse dwarf-like galaxies 
in terms of their 
stellar mass and DM content, with comparable colours to those of dwarf galaxies in the same range of luminosity. 
This result, however, suffers from the large uncertainties coming from the $N_{GCs}$ estimates. 
In particular, due to the low number of tracers and large 
contamination from fore- and background objects, the DM estimate from GCs is quite uncertain. Follow-up studies, which include both
deep multi-band imaging data, preferably including near-IR bands, coupled with a more detailed analysis 
over the entire area surveyed (which is in progress), and integral field spectroscopy, 
will substantially help in reducing such uncertainty.
Future systematic searches for UDGs and their 
analysis in this cluster, will be fundamental to check whether DM-dominated systems are also present and,
therefore, to put further observational constraints on 
the existence of two distinct formation channels for this class of galaxies.}


Taking into account the virial mass of the Hydra I cluster 
\citep[$2\times 10^{14}$~M$_{\odot}$,][]{Girardi1998},
from the abundance-halo mass relation \citep{Janssens2019} 
we expect up to 100 UDGs inside the cluster virial radius. 
A follow up study, where the UDG detection is not carried out visually, but using automated tools already tested on the Fornax galaxy cluster \citep{Venhola2019}, is in progress.

This work is part of a large program to study the internal structure, formation history, evolution and DM fraction 
in UDGs across different environments 
identified from the deep, wide-field imaging data of VEGAS. By using automatic LSB detection tools on the entire VEGAS sample, which includes about 30 groups and 10 clusters of galaxies  
for a total covered area of more than $\sim 100$ deg$^2$, we aim at identifying and studying 
a large number ($\sim$1300) of UDGs. The new sample will almost 
double the number of studied UDGs and will have a 
legacy value for future follow-up imaging and spectroscopic observations.

 \begin{acknowledgements}
This work is based on visitor mode observations taken at the ESO 
La Silla Paranal Observatory within the VST Guaranteed Time Observations, Programme ID: 099.B-0560(A). {  We thank the anonymous referee for his/her useful suggestions that helped to improve the paper.}
EI acknowledge financial support from the ESO during the science visit at the Garching HQ, from 
September 1st, 2019 up to August 2020.
MS and EI acknowledge finacial support from the VST
project (P.I. P. Schipani). Authors wish to thank ESO for the
financial contribution given for the visitor mode runs at the ESO La Silla Paranal Observatory.
GD acknowledges support from CONICYT project Basal AFB-170002.

\end{acknowledgements}

 \bibliographystyle{aa.bst}
  \bibliography{Hydra}


\begin{table*}
\setlength{\tabcolsep}{2pt}
\small 
\caption{Parameters of the UDG candidates in the Hydra I cluster.} 
\label{tab:UDGsample}
\vspace{13pt}
\begin{tabular}{lcccccccccc}
\hline\hline
Object & R.A. & DEC & $M_r$ & $g-r$ & M/L & M$_*$ & $\mu_{0}$ &  $R_{e}$ & $n$& $q$ \\ 
 &J2000&J2000 & [mag] & [mag] &  & [$10^{8}$~M$_\odot]$ &[mag/arcsec$^{2}$] & [kpc]& & \\
\hline \vspace{-7pt}\\
Hydra~I-UDG 1 &10:37:54.12 & -27:09:37.50& -15.48 $\pm$ 0.07 & 0.40 $\pm$ 0.09 & 0.90 &1.12 & 24.20$\pm$0.10 & 1.75$\pm$0.12 & 0.621$\pm$0.005& 0.766$\pm$0.003 \\ 
Hydra~I-UDG 2 & 10:37:34.89 & -27:10:29.94& -14.27 $\pm$ 0.05 & 0.53 $\pm$ 0.2 & 1.35 & 0.55 & 24.97$\pm$0.08 & 1.55$\pm$0.12 & 0.72$\pm$0.01& 0.877$\pm$0.007 \\
Hydra~I-UDG 3 & 10:36:58.63 & -27:08:10.21 & -14.7 $\pm$ 0.2 & 0.75 $\pm$ 0.3 & 2.71 & 1.65 &25.2$\pm$0.2 & 1.88$\pm$0.12 & 0.572$\pm$0.013& 0.79$\pm$0.01 \\
Hydra~I-UDG 4 & 10:37:02.64 & -27:12:15.01 & -16.03 $\pm$ 0.04 & 0.95 $\pm$ 0.10 & 5.11 & 10.6 &24.86$\pm$0.05 & 2.64$\pm$0.12 & 0.597$\pm$0.006& 0.748$\pm$0.003 \\
Hydra~I-UDG 5 & 10:36:07.68 & -27:19:03.26 & -14.66 $\pm$ 0.09 & 0.65 $\pm$ 0.14 & 1.98 & 1.16 &23.7$\pm$0.3 & 1.42$\pm$0.12 & 0.89$\pm$0.01& 0.582$\pm$0.004 \\
Hydra~I-UDG 6 & 10:36:35.80 & -27:19:36.12 & -14.38 $\pm$ 0.08 & 0.32 $\pm$ 0.2 & 0.70 & 0.32 &24.08$\pm$0.13 & 1.37$\pm$0.12 & 0.706$\pm$0.006& 0.629$\pm$0.002 \\
Hydra~I-UDG 7 & 10:36:37.16 & -27:22:54.93& -13.72 $\pm$ 0.13 & 0.6 $\pm$ 0.3 & 1.98 & 0.49 &24.37$\pm$0.4 & 1.66$\pm$0.12 & 1.32$\pm$0.03& 0.73$\pm$0.01 \\
Hydra~I-UDG 8 & 10:38:14.59 & -27:24:27.07 & -14.87 $\pm$ 0.07 & 0.34 $\pm$ 0.13 & 0.74 & 0.53 &23.22$\pm$0.6 & 1.40$\pm$0.12 & 0.98$\pm$0.02& 0.578$\pm$0.005 \\
Hydra~I-UDG 9 & 10:37:22.85 & -27:36:02.80 & -15.16 $\pm$ 0.12 & 0.6 $\pm$ 0.2 & 1.92 & 1.78 & 24.2$\pm$0.2 & 3.46$\pm$0.12 & 1.38$\pm$0.01& 0.54$\pm$0.01 \\
Hydra~I-UDG 10 & 10:35:27.32 & -27:33:03.86 & -13.89 $\pm$ 0.08 & 0.4 $\pm$ 0.2 & 0.90 & 0.26 & 24.33$\pm$0.3 & 2.29$\pm$0.10 & 1.53$\pm$0.02& 0.90$\pm$0.01 \\
Hydra~I-UDG 11 & 10:35:04.16 & -27:26:17.26 & -14.75 $\pm$ 0.07 & 0.43 $\pm$ 0.11 & 0.99 & 0.63 &24.36$\pm$0.13 & 1.66$\pm$0.12 & 0.80$\pm$0.01& 0.92$\pm$0.01 \\
Hydra~I-UDG 12 & 10:34:59.55 & -27:25:37.95 & -14.3 $\pm$ 0.2 & 0.8 $\pm$ 0.4 & 2.8 & 1.19 &25.1$\pm$0.2 & 1.64$\pm$0.12 & 0.67$\pm$0.02& 0.72$\pm$0.01 \\
\hline
\end{tabular}
\tablefoot{Column 1 reports the name of the UDG candidate. In columns 2 and 3 we list the coordinates of the UDGs. In columns 4 and 5 are reported the total $r$-band magnitude and the average $g-r$ colour. 
Columns 6 and 7 give the stellar mass-to-light ratio and stellar mass, respectively, derived in the $r$ band.
Columns 8 to 11  list the structural parameters derived from the 2D fit in the $g$ band: the central surface brightness, the effective radius in kpc, the $n$ exponent of the S{\'e}rsic law and the axial ratio, respectively. Magnitudes and colours are corrected for Galactic extinction using values from \citet{Schlegel98}.}
\end{table*}

\begin{table*}
\setlength{\tabcolsep}{2pt}
\small 
\caption{LSB candidates in the Hydra I cluster.} 
\label{tab:LSBsample}
\vspace{10pt}
\begin{tabular}{lcccccccc}
\hline\hline
Object & R.A. & DEC & $M_r$ & $g-r$ & $\mu_{0}$ &  $R_{e}$ & $n$& $q$ \\ 
      &J2000 & J2000 & [mag] & [mag] & [mag/arcsec$^{2}$] & [kpc]& & \\
\hline \vspace{-7pt}\\
Hydra~I-LSB 1 &10:35:17.540 & -27:10:51.96& -13.31 $\pm$ 0.14 & 1.1 $\pm$ 0.4 &  24.2$\pm$0.3 & 0.86$\pm$0.12 & 1.05$\pm$0.02& 0.713$\pm$0.009 \\ 
Hydra~I-LSB 2 & 10:34:53.099 & -27:12:00.06& -15.12 $\pm$ 0.04 & 0.69 $\pm$ 0.09 &  23.57$\pm$0.13 & 1.08$\pm$0.12 & 0.69$\pm$0.004& 0.903$\pm$0.003 \\
Hydra~I-LSB 3 & 10:34:37.181 & -27:12:41.14 & -14.55 $\pm$ 0.09 & 0.34 $\pm$ 0.15 & 23.73$\pm$0.14 & 0.997$\pm$0.12 & 0.660$\pm$0.004& 0.979$\pm$0.003 \\
Hydra~I-LSB 4 & 10:35:37.232 & -27:14:50.14 & -14.25 $\pm$ 0.05 & 0.58 $\pm$ 0.12 & 23.5$\pm$0.2 & 0.62$\pm$0.12 & 0.580$\pm$0.005& 0.925$\pm$0.003 \\
Hydra~I-LSB 5 & 10:37:54.147 & -27:16:15.31 & -13.64 $\pm$ 0.11 & 0.4 $\pm$ 0.2 & 25.65$\pm$0.14 & 1.29$\pm$0.12 & 0.58$\pm$0.02& 0.858$\pm$0.013 \\
Hydra~I-LSB 6 & 10:38:05.430 & -27:15:46.89 & -14.08 $\pm$ 0.06 & 0.60 $\pm$ 0.11 & 24.0$\pm$0.2 & 1.02$\pm$0.12 & 0.882$\pm$0.012& 0.763$\pm$0.006 \\
Hydra~I-LSB 7 & 10:37:43.555 & -27:20:44.73& -13.19 $\pm$ 0.05 & 0.55 $\pm$ 0.11 & 25.00$\pm$0.15 & 1.21$\pm$0.12 & 0.63$\pm$0.02& 0.55$\pm$0.01 \\
Hydra~I-LSB 8 & 10:35:42.806 & -27:20:09.61 & -12.35 $\pm$ 0.13 & 0.4 $\pm$ 0.2 & 25.60$\pm$0.13 & 0.92$\pm$0.13 & 0.78$\pm$0.03& 0.94$\pm$0.02 \\
Hydra~I-LSB 9 & 10:36:16.856 & -27:36:50.83 & -13.01 $\pm$ 0.09 & 0.7 $\pm$ 0.2 & 25.12$\pm$0.12 & 1.00$\pm$0.12 & 0.73$\pm$0.02& 0.63$\pm$0.01 \\
Hydra~I-LSB 10 & 10:35:03.551 & -27:25:50.07 & -14.46 $\pm$ 0.10 & 0.9 $\pm$ 0.3 & 23.6$\pm$0.2 & 0.98$\pm$0.12 & 0.609$\pm$0.006& 0.675$\pm$0.003 \\
Hydra~I-LSB 11 & 10:34:56.212 & -27:32:57.79 & -15.32 $\pm$ 0.07 & 0.29 $\pm$ 0.09 & 22.86$\pm$0.15 & 1.14$\pm$0.12 & 0.765$\pm$0.003& 0.834$\pm$0.001 \\
Hydra~I-LSB 12 & 10:38:03.959 & -27:45:28.18 & -14.40 $\pm$ 0.05 & 0.9 $\pm$ 0.2 & 24.3$\pm$0.2 & 1.28$\pm$0.12 & 0.627$\pm$0.011& 0.484$\pm$0.004 \\
Hydra~I-LSB 13 & 10:37:19.034 & -27:42:29.29 & -13.95 $\pm$ 0.07 & 0.62 $\pm$ 0.14 & 24.53$\pm$0.14 & 1.08$\pm$0.12 & 0.56$\pm$0.01& 0.653$\pm$0.006 \\
Hydra~I-LSB 14 & 10:36:41.231 & -27:48:20.04 & -13.85 $\pm$ 0.08 & 1.0 $\pm$ 0.4 & 24.0$\pm$0.3 & 1.03$\pm$0.12 & 0.81$\pm$0.01& 0.546$\pm$0.005 \\
Hydra~I-LSB 15 & 10:34:42.961 & -27:40:14.42 & -14.96 $\pm$ 0.10 & 0.6 $\pm$ 0.2 & 23.1$\pm$0.2 & 0.89$\pm$0.12 & 0.795$\pm$0.004& 0.920$\pm$0.003 \\
\hline
\end{tabular}
\tablefoot{Column 1 reports the name of the LSB candidate and in columns 2 and 3 we list the coordinates. 
In columns 4 and 5 are reported the total $r$-band magnitude and the average $g-r$ colour. 
Columns 6 and 9  list the structural parameters derived from the 2D fit in the $g$ band: the central surface brightness, the effective radius in kpc, the $n$ exponent of the S{\'e}rsic law and the axial ratio, respectively. Magnitudes and colours are corrected for Galactic extinction using values from \citet{Schlegel98}.}
\end{table*}

\begin{table*}
\setlength{\tabcolsep}{2pt}
\caption{Globular cluster candidates and halo mass estimates for the UDGs.} 
\label{tab:GCs}
\vspace{13pt}
\begin{tabular}{lcccccc}
\hline\hline
ID & $N_{GC}$  & $N_{GC}$  & $M_h$ & $M/L_V$ & $S_N$ & Nuclear star cluster candidates\\ 
    & ( $\leq 1.5R_e$) & ({  $\leq 5R_e$}) &  $10^{10}$~M$_\odot$ & $10^3$ & &\\
\hline \vspace{-7pt}\\
Hydra~I-UDG 1 &  0$\pm$1 & 0 $\pm$2  &  \nodata   & \nodata & \nodata  & \nodata \\ 
Hydra~I-UDG 2 &  {  7 $\pm$3 }& {  3 $\pm$2 } & {  1.4 $\pm$ 0.9 }& {  0.43} &{  7$\pm$5 }&  $m_g\sim24.2$, and $\sim25.2$, both within $R_{proj}=1$~arcsec  \\
Hydra~I-UDG 3 & {  15$\pm$6 }& { 6$\pm$2}  &{   3$\pm$1}   & { 0.96  }   & { 11$\pm$4 }& $m_g\sim24.2$ at $R_{proj}\leq 1$~arcsec, $m_g\sim22.5$ at $R_{proj}\leq 7$~arcsec\\
Hydra~I-UDG 4 & {  2$\pm$1 }& { 3$\pm$4}  &{   1$\pm$1}   & { 0.034}    &{  2$\pm$2  }& \nodata \\
Hydra~I-UDG 5 & {  0 $\pm$1} & { 0 $\pm$1}  &{  \nodata}& {  \nodata} &{  \nodata}& \nodata \\
Hydra~I-UDG 6 & {  0 $\pm$1} & { 0 $\pm$1}&  \nodata   & \nodata & \nodata  & \nodata\\
Hydra~I-UDG 7 & {  3 $\pm$1 }& { 3$\pm$2}  &{   1.4$\pm$0.9}   &{  0.25 }    & { 13$\pm$8} & $m_g\sim26.2$ at $R_{proj}\leq 1\arcsec$  \\
Hydra~I-UDG 8 & {  0 $\pm$1} & { 0 $\pm$0 }  &  \nodata   & \nodata & \nodata  &  \nodata \\
Hydra~I-UDG 9 & {  7$\pm$3 }& { 10$\pm$8} &{   5$\pm$4} & { 0.40} &{  11$\pm$9}&  \nodata \\
Hydra~I-UDG 10 & {  0$\pm$1 }& { 0 $\pm$3}  &{   \nodata }  & {  \nodata}&{  \nodata}& 
Two partially blended sources at $R_{proj}\leq 1$~arcsec \\
Hydra~I-UDG 11 & {  7$\pm$3} & { 5$\pm$2}  &{   2.4$\pm$0.9}   & { 0.35}& { 7$\pm$3 }& Diffuse nuclear source \\
Hydra~I-UDG 12 & {  0$\pm$1} &{ 0$\pm$1  }&  \nodata   & \nodata & \nodata  &  \nodata\\
\hline
\end{tabular}
\tablefoot{Column 1 reports the name of the UDG candidate. Columns
  2 and 3  give the total number of GC candidates  inside $R=1.5R_e$ and
  {  $5R_e$}, respectively. Columns 4 and 5 report the halo mass
  and the $V$-band halo mass-to-light ratio,
  respectively.  Column 6 lists the GC specific frequency {  derived from the total number of GCs inside $5R_e$}. Column 7
  provides some details on the presence of nuclear star clusters (NSC); if present, the $g$ magnitude and the projected galactocentric distance from the galaxy core, $R_{proj}$, are reported.}
\end{table*}

\end{document}